\documentclass{andp2012}
\usepackage[english]{babel}
\title{Transient excitonic states in optically-pumped Dirac materials: overview of recent work}
\author[A. Pertsova]{A. Pertsova\inst{1}\footnote{Corresponding author\quad E-mail:~\textsf{anna.pertsova@su.se}}}
\author[A.\,V. Balatsky]{A.\,V. Balatsky\inst{1,2}}
\address[1]{Nordita, Roslagstullsbacken 23, SE-106 91 Stockholm, Sweden}
\address[2]{Department of Physics, University of Connecticut,  Storrs, CT 06269, USA}
\begin{abstract}
  Driven and non-equilibrium quantum states of matter have attracted growing interest in both theoretical and experimental studies
   in condensed matter physics.
  We review recent progress in realizing transient collective states in driven or pumped Dirac materials (DMs).
  In particular, we focus on optically-pumped DMs which have been theoretically proposed as a promising platform for
  observation of a transient excitonic instability.
  Optical pumping combined with the linear (Dirac) dispersion of the electronic spectrum offers a knob for tuning the
  effective interaction between the photoexcited electrons and holes, and thus provides a way of
  reducing the critical coupling for excitonic instability.
  As a result, a transient excitonic condensate could be achieved in a
  pumped DM while it is not feasible in equilibrium. We provide a 
  unifying theoretical framework for describing transient collective states in two- and three-
  dimensional DMs. 
  We  describe experimental signatures of the transient excitonic state and
   summarize numerical estimates of the magnitude of the effect, namely the size of the 
   dynamically-induced excitonic gaps and the values of the critical temperatures  
   for several specific systems. We also 
 discuss general guidelines for identifying promising material candidates.
 Finally, we comment recent experimental efforts in
 realizing transient excitonic condensate in pumped DMs
 and outline outstanding issues and possible future directions.
\end{abstract}
\shortabstract

\begin{document}
\maketitle

\section{Introduction}
The last decade has seen an increased interest in generating and controlling dynamical and non-equilibrium states of matter.
Such states can be induced by interaction with a time-dependent drive such as electromagnetic-fields or coherent lattice vibrations.
Prominent recent examples include novel dynamical states in periodically-driven systems, such as Floquet topological insulators
~\cite{Lindner2011,Sentef2015,Claassen2016,Zou_prb2016,Liu2016,Hubener2017} and
time crystals~\cite{Wilczek_prl2013,Khemani_prl2016,Yao_prl2017,Choi2017},
phonon-driven Floquet states~\cite{Hubener2018}, light-induced superconductivity~\cite{Fausti189,Mitrano2016}, and transient, or non-equilibrium, 
exciton
condensate in
 pumped semimetals~\cite{Triola_prb2017,Pertsova_prb2018} and semiconductors~\cite{Szymaska_prl2006,versteegh2012pumpedZnO,Perfetto2019}.
 Another example is intrinsically dynamical orders such as odd-frequency, or Berezinskii, superconducting pairing~\cite{Berezinskii1974, Oddf2017}, 
 which
 involves correlations non-local in time
and naturally manifests in the time domain, see also related paper in this volume \cite{BalatskyQTO2019}.

Progress is investigating the dynamical states is fueled by the development of time-sensitive pump-probe   
techniques such as 
time- and angle-resolved 
 photoemission spectroscopy (trARPES) and
 time-resolved terraherz (THz) spectroscopy, which are able to probe the non-equilibrium dynamics of electronic states at ultrafast 
(subpicosecond) time-scales.
Ultrafast spectroscopy is currently used to study quantum materials, i.e. materials where strong quantum
correlations lead to unusual emergent properties such as unconventional orders and topology~\cite{Orenstein2012, QM2017}.
Such studies may provide insights into the physics of
quantum materials and reveal novel dynamically-induced states that are inaccessible in equilibrium.

In this contribution, we discuss recent efforts towards the realization of transient collective states in a particular class of quantum materials,
known as Dirac materials (DMs)~\cite{wehling2014dirac}.
DMs are characterized by nodes and a linear, Dirac-like, dispersion of the quasiparticle spectrum. This is a growing class of materials
encompassing high-temperature $d$-wave superconductors~\cite{balatsky_rmp2006},
superfluid $^3$He~\cite{volovik_he3}, graphene~\cite{neto2009electronic}, topological insulators (TI)~\cite{hasan_rmp2010,qi_rmp2011},
 Dirac (DSM)~\cite{neupane2014dirac} and
Weyl semimetals (WSM)~\cite{huang2015weyl,xu2015weyl}, Dirac nodal line semimetals~\cite{burkov_prb2011_linenodes,sun_prb2017_linenodes} and
 bosonic DMs~\cite{fransson_prb2016,banerjee_prb2016}.

Due to the great potential of DMs for high-performance optoelectonic devices~\cite{otsuji2012graphene,Wang_nanolett2017_DSM_photodetector},
non-equilibrium dynamics
of DMs has
emerged as an important research topic. Many studies have focused on the interplay between light and Dirac states~\cite{Dolcini_prb2016_1Dpump,
Wang_prl2016_light_dm,sanchez-barriga_prx2014_light_dm} and, in particular, on
 optical generation and control of spin-polarized currents on TI surfaces~\cite{Kuroda_prl2016_light_dm}.
 Ultrafast spectroscopies have been used extensively to study electron dynamic in 2D DMs, e.g.
 graphene~\cite{george2008ultrafast,gilbertson2011tracing,li2012femtosecond,aguilar2015time,gierz2013graphene,johannsen2013graphene,
 ulstrup2014bilayer,Johannsen2015Graphene,gierz2015graphene}
 and 3D TIs~\cite{zhu2015ultrafast,neupane2015gigantic}.
  Ultrafast dynamics of the recently discovered 3D DMs, i.e. DSM and WSM, is a less explored topic but
 pump-probe studies similar to 2D DMs have recently appeared in the literature~\cite{Manzoni_prl2015_ultrafast_DSM,Ishida_prb2016_relaxation_DSM,
 Jadidi_THz_pump-probe_SM,Ma_natphys2017_pumping_WSM}.
 In particular, relaxation dynamics was studied in TaAs~\cite{Jadidi_THz_pump-probe_SM, Weber_japl2017}, NbP~\cite{Weber_japl2017}
 and NbAs~\cite{Weber_apl2018} WSMs, and
 in Cd$_3$As$_2$ DSM~\cite{Lu_prb2017_ultrafast_DSM}.

An important result of the pump-probe studies in DMs is
the possibility of creating a broadband
population inversion~\cite{li2012femtosecond,Sumida_prb2019}. In the context of 
optical pumping, population inversion is a situation when after initial rapid 
thermalization two independent Fermi-Dirac distributions are established, with  
 distinct chemical potentials for the non-equilibrium electron and hole populations 
 (see Fig.~\ref{fig1}). 
Population inversion 
is crutial for realizating broadband THz lasing in DMs~\cite{otsuji2012graphene}. 
According to the most optimistic experimental reports, 
 population inversion in graphene can be sustained on the timescale of $100-200$~fs
~\cite{li2012femtosecond,gierz2013graphene,johannsen2013graphene,ulstrup2014bilayer,Johannsen2015Graphene}.

More recently, long-lived optically-excited states have been demonstrated in 3D TIs with lifetimes ranging
from few ps to hubdreds of ps, e.g.  $\tau\approx 3$~ps in Sb$_2$Te$_3$~\cite{zhu2015ultrafast} and 
$\tau\approx 400$~ps in (Sb$_{1-x}$Bi$_x$)$_2$Te$_3$, where the chemical potential is positioned 
 inside the bulk insulating gap~\cite{sumida2017}.
In particular, a recent time-resolved ARPES study provided evidence of the population inversion in the $p$-type
3D TI (Sb$_{0.73}$Bi$_{0.27}$)$_2$Te$_3$
on the time scale of $10$~ps~\cite{Sumida_prb2019}, which is two orders of magnitude larger than in graphene.
 The prolonged lifetime of population inversion in 
 3D TIs with chemical potential close to the Dirac node is 
  attributed to the relaxation bottleneck originating from the reduced phase space near the node.
  
Although it is not clear at present whether
  population inversion is observed in 3D DMs, the current understanding is that
  the ultrafast carrier dynamics in 3D DMs and graphene is qualitatively similar. The overall lifetime of
photoexcited carriers in 3D DMs is of the order of few ps~\cite{Weber_japl2017, Weber_apl2018}.
 The possibility of achieving an inverted population of electrons and holes by optical pumping, 
 makes optically excited DMs a promising system for realizing a \textit{transient excitonic instability}
 ~\cite{Triola_prb2017,Pertsova_prb2018}.

By definition,  excitonic instability in equilibrium occurs
when the exciton binding energy exceeds the (positive or negative) band
gap of the material. This can be realized in narrow gap semiconductors, semimetals or metals with overlapping bands.
In these systems, at sufficiently low energies, the Coulomb attraction between electrons and holes
residing in the conduction and valence band respectively, leads to a new collective ground state known as an \textit{excitonic insulator}, or
 the electron-hole Bardeen-Copper-Schrieffer (BCS) state,
 i.e. a condensate of electron-hole Cooper pairs~\cite{Keldysh1964,jerome1967excitonic,halperin1968possible}.
This should be distinguished from a Bose-Einstein condensate (BEC) of excitons, or bound states of a single electron-hole 
pair~\cite{Blatt_prb1962,Keldysh1968}.
 A transition from an exciton BEC to a BCS state has been predicted for increasing densities of electrons and holes at low
 temperatures~\cite{Comte1982}.

Many systems have been suggested as suitable for excitonic condensation. These include systems of ``direct'' excitons, e.g.
narrow-gap semiconductors where paired electrons and holes reside in the conduction and valence bands of the same
material~\cite{halperin1968possible, jerome1967excitonic}.
Signatures of the excitonic insulator phase have been reported in semiconductors TmSe$_{0.45}$Te$_{0.55}$~\cite{Bucher_prl1991},
Ta$_2$NiSe$_5$~\cite{Wakisaka_prl2009} and
1$T$-T$_1$Se$_2$~\cite{Monney_PRL2011}. Another example is condensation of ``indirect'' excitons realizable in bilayer systems, 
where electrons and holes are
spatially separated~\cite{lozovik1976superconductivity,Shevchenko1976,zhu1995exciton,Conti_PRB1998,Pieri_prb2007}.
 Evidence of BEC of indirect excitons has been reported in electron-hole bilayers found in semiconductor quantum wells~\cite{Butov_prl1994,
 butov2002excitons,High_nature2012,High_nanolett2012}. However, the most clear signatures of the exciton BEC have been observed in quantum 
 Hall electron-electron
 bilayers in strong magnetic field~\cite{Kelogg_prl2004,tatuc_prl2004,eisenstein2004bose,Nandi2012}.
 Gated graphene bilayers~\cite{Zhang_prb2008,Min_prb2008,Kharitonov_prb2008} and TI thin films~\cite{Seradjeh_prl2009,
 Tilahun_prl2011} have also been proposed as promising platforms.
 Recently, a bilayer exciton BEC has been proposed in quasi-one-dimensional systems with
 a single particle hybridization between electron and hole populations~\cite{Kantian_prl2017}.

In principle, electron-hole pair formation is also possible in highly excited semiconductors,
where electron and hole populations are created by optical pumping in the conduction and valence bands respectively.
In this case the
possible excitonic insulator phase or the exciton BEC are intrinsically \textit{non-equilibrium phenomena}
 due to finite lifetime of electron and hole populations. As commonly accepted in the literature,
 in this work we refer to these states collectively as a transient excitonic condensate.
Such systems have been studied extensively in theoretical works~\cite{Haug_prb1988,Ostreich1993,
Littlewood_2004,Szymaska_prl2006,Hanai2016,Hanai_prb2017,Hanai_prb2018,Perfetto2019}.
Experimentally, clear signatures of a transient excitonic condensate
 are yet to be observed. There are indications of possible ordered exciton state in CuO$_2$~\cite{Yoshioka2011}.
Also, signs of preformed (uncondensed) electron-hole pairs in measurements of 
stimulated emission in highly excited ZnO were reported~\cite{versteegh2012pumpedZnO}
More recently, pumped black phosphorus  was suggested as
 a promising candidate for excitonic condensation, with lifetime of carriers excited across the band gap exceeding
 $400$~ps~\cite{Nurmamat2018}.

In a different setting,
  semiconductor-based   microcavities provide a convenient platform for
  condensation. In these systems photons are confined and strongly coupled to electronic excitations,
  leading to the creation of exciton polaritons. Condensation is favorable due to the very small mass of the
polaritons and BEC of exciton polaritons has indeed been observed in these systems~\cite{Kasprzak2006,Balili,Kim2011}.
Among other work on non-equilibrium many-body states, we mention Floquet exciton condensation in bilayer graphene~\cite{zhang2015floquet}
and photo-induced superconducting states in semiconductors
under optical driving~\cite{goldstein2015photoinduced}.

The proposal for realization of the transient excitonic condensate in optically-excited DMs differs from previous work
in several important aspects. Firstly, we consider a semimetal with a very special energy dispersion. 
The linear (Dirac) dispersion relation in DMs results in the strongly energy-dependent density of states (DOS):  
$\mathcal{N}(E)\propto E$ or $\mathcal{N}(E)\propto E^2$ in 2D and 3D DMs, respectively. 
 The presence of nodes in the low-energy spectrum implies vanishing DOS at the nodes,  
 which leads to a critical coupling for many-body 
instabilities~\cite{kotov2012electron}.

The strength of electron-electron interactions is typically characterized 
by a material-specific parameter, the so called dimensionless coupling constant $\alpha$, which 
is expressed as a ratio between the Coulomb energy and the kinetic energy. 
 In the case of DMs, it is given by  $\alpha=e^2/\hbar\varepsilon v$, where $\varepsilon$ is the 
material-dependent dielectric constant and $v$ is the Fermi velocity of the linearly-dispersing states. 
 Critical coupling, $\alpha_c$, is defined as a value of the dimensionless coupling constant such that for $\alpha<\alpha_c$, 
  the system is not 
significantly affected by interactions and the spectrum remains gapless, while for $\alpha>\alpha_c$, the spectrum aquires an energy gap
 due to interactions. The gapped phase can be referred to as the excitonic insulator. 
 Many theoretical studies suggest that in the case of 2D DM (graphene), $\alpha_\mathrm{c}\approx 1$
 ~\cite{kotov2012electron,drut2009graphene,gamayun2010gap,gamayun2009supercritical}.
  The excitonic insulator phase has not yet been observed in DMs. Experimental studies 
  on suspended graphene, where $\alpha$ is expected to be larger than $1$ ($\alpha\approx 2.2$),
   show that the only effect of electron-electron interactions is the logariphmic renormalization of 
   the Fermi velocity close to the node~\cite{elias2011dirac}.
We propose to use optical pumping to generate excitonic gaps in DMs.

Secondly, optical pumping in DMs 
 allows for a tunable enhancement of the DOS of the inverted electron and hole populations, and, thus,
 offers a tuning knob for the effective Coulomb interaction. 
 This is a unique feature of DMs, which arises from the stongly energy-dependent DOS. 
 For instance, it is not achievable in metals or semiconductors in 2D, since in this case 
the DOS is constant at low energies. 
Finally, an important signature of the transient excitonic state are the energy gaps that 
open up in the quasiparticle spectrum at the non-equilibrium chemical potentials for electrons and holes. 
Although the gaps have a transient nature,
our proposal, if realized, offers  an alternative way of generating band gaps in DMs, which 
is highly desirable for application in optoelectronics.

We develop a unified theoretical description of 
trasient excitonic states in optically-pumped 2D and 3D DMs. 
The theory is based on a low-energy effective model for DMs. 
As a first step, we consider excitonic instabilities in 2D and 3D DMs with population 
inversion by using the mean-field BCS theory, applied to a system with two types of carriers (electrons and holes).  
 Screening effects are important in pumped DMs, where the non-equilibrium chemical potentials are located far from the 
 Dirac node, and are included in the Thomas-Fermi approximation. By substituting material parameters in the model, 
 we obtain numerical estimates for the values of critical temperature ($T_c$) and excitonic gap ($\Delta$) for 
 realistic materials. The relaxation of
the transient excitonic state towards equilibrium is studied using a dynamical model based on semiconductor Bloch equations.

We predict the largest effect, i.e. a gap of the order of $10$~meV and a critical temperature of $70$~K in 
 prestine suspended graphene, in which pumping is realized selectively on a single valley. 
 Large gaps and critical temperatures should enable detection of the transient excitonic 
 states in gaphene by trARPES; however, the relatively short lifetime ($100-200$~fs) is likely to make the observation challenging.
 Another promising 2D system is a 3D TI with a single surface Dirac cone such as Bi$_2$Se$_3$ and related materials. Although 
 the estimated values for the gap and $T_c$ are typically smaller in 3D TIs compared to graphene due to smaller $\alpha$, 
 the lifetime of the population inversion is at least an order of magnitude larger. Further tuning material parameters such as 
 the bulk gap and the dimensionless coupling constant, can lead to a much larger effect in 3D TIs, namely $T_c$ of the order of 
 $100$~K and gaps of $10$s of meV.  In the case of DSM and WSM, severe screening effects due to large 
 Dirac cone degeneracy in existing materials, limits the estimated gaps sizes to $1$~meV and critical temperature to few K. 
  By using the recipe for enhancement of the excitonic gap and $T_c$, similarly to the case of 2D DMs, large values, comparable to 
  those predicted for graphene, could be also obtained in future 3D DMs.

The rest of the paper is organized as follows. In Section~\ref{model}, we present the mean-field theory of excitonic pairing
in pumped DMs. In Section~\ref{exp:sig}, we discuss spectroscopic signatures and calculate the order 
 parameter as a function of  temperature, non-equilibrium chemical potentials and dimensionless coupling constant. 
In Section~\ref{exp:mater}, we summarize the numerical estimates 
for several examples of realistic and hypothetical 2D and 3D DMs and suggest
experimental setups in which the effect could be observed.
 We also derive general criteria for achieving large excitonic gaps and $T_c$ 
 and suggest routes for search of promising material candidates. Finally,
 in Section~\ref{concl} we offer some conclusions and outlook.

\section{General theory of pumped DMs}\label{model}
\subsection{Pumping scheme}\label{pump_scheme}
\begin{figure*}[ht!]
\centering
\includegraphics[width=0.9\linewidth,clip=tue]{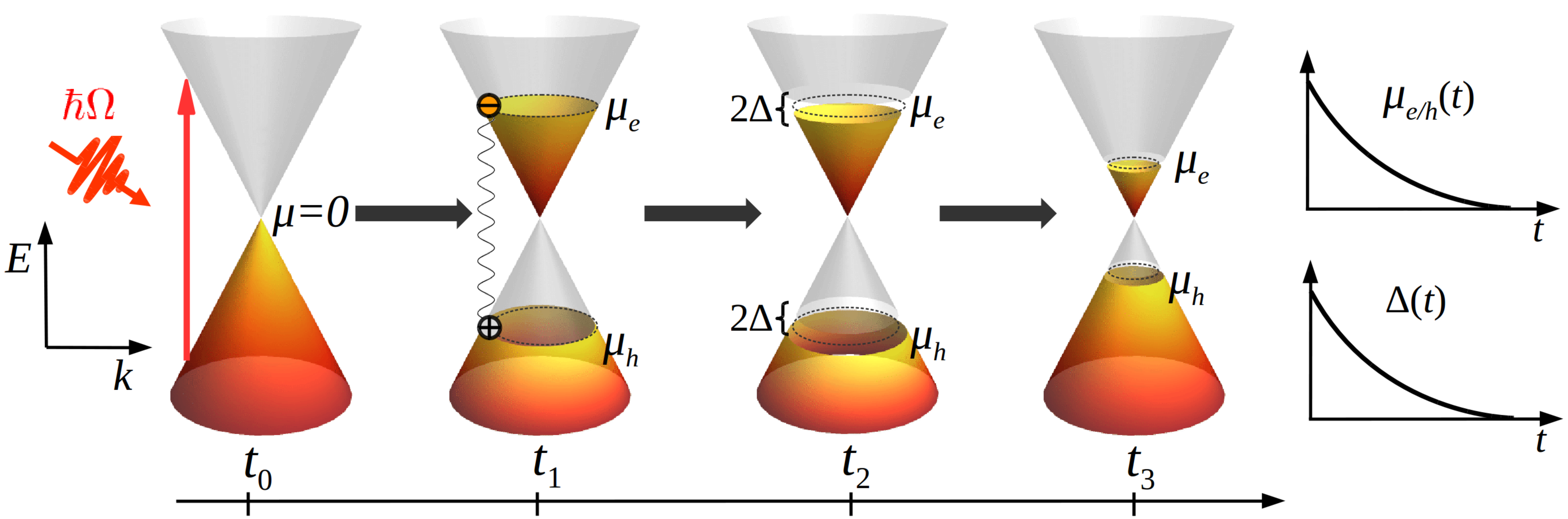}
\caption{Transient excitonic instability is 
 a pumped DM with a population inversion.  
 Before the pump, at time $t<t_0$, the system is in equilibrium and is described by 
 a single chemical potential $\mu=0$; 
at $t=t_0$ electrons are excited from the valence
band to the conduction band; after thermalization time $t=t_1$, 
 a population inversion is established, i.e. photoexcited electrons and holes 
  form two distinct Fermi-Dirac distributions
with chemical potentials $\mu_{\mathrm{e}}$ and $\mu_{\mathrm{h}}$, respectively; 
 pairing between finite populations of electrons and holes leads to 
 formation of a transient excitonic state at $t=t_2$,  
characterized by gap oppenings at non-equilibrium chemical potentials;
 due to a finite lifetime of the population inversion, 
 transient excitonic states decay towards equilibrium ($t=t_3$).
 Panels on the right show schematically the relaxation
 of the chemical potentials $\mu_{\mathrm{e}}$ and the order parameter $\Delta$
 towards equilibrium. For illustration,  we consider a
 single Dirac cone and only show the 2D energy dispersion of the Dirac states 
 (for a 3D DM, we assume a 2D projection of the 3D Dirac states). 
 Empty Dirac states are indicated in white, filled states in
yellow.}
\label{fig1}
\end{figure*}
Pump-probe experiments suggest that population inversion can be achieved in optically-excited DMs. The pumping scheme
may differ depending on the material. In graphene, Dirac states extend in
the energy window of approximately $1$~eV;
therefore, an optical excitation promotes electrons from the occupied
states in the lower Dirac cone into the unoccupied states in the upper Dirac cone (this situation is depicted in Fig.~\ref{fig1}).
In a typical 3D TI like Bi$_2$Se$_3$, with a band gap of few hundred meV, a pump pulse excites electrons from the bulk valence band into the bulk
conduction band. The excited carriers then cascade into the unoccupied states in the upper Dirac cone. Vanishing phase-space
for excitations near at the Dirac point acts as a bottleneck for relaxation processes.
Pumping on 3D DMs occurs in a similar fashion since the energy extent of the 3D Dirac states in typically small
compared to the energy of the pump pulse. However, recently dynamics of photoexcited carriers
 was probed in 3D DMs by direct excitation with low-energy photons~\cite{Weber_japl2017}, similarly to graphene.
Based on these observations, a schematic diagram of a pumped DM is drawn in Fig.~\ref{fig1}.

In equilibrium ($t<t_0$), we consider a DM with a chemical potential at the Dirac point ($\mu=0$). 
 The pump pulse is switched on at $t=t_0$ and promotes electrons from the occupied states below  to 
 the empty states above the Dirac point. We assume that after rapid thermalization, the population inversion 
 is established at $t=t_2$ and relaxes towards a single Fermi-Dirac distribution on the timescale $t_3$ (in general, 
 at a higher temperature than in equilibrium).  
 The lifetime is of the order of $100$~fs in graphene~\cite{li2012femtosecond}; however, we should note that 
  there are other trARPES reports that do not show any population inversion in graphene within the temporal 
  resolution of the experiment~\cite{johannsen2013graphene}.
In 3D TIs, the lifetime is extended to at least several ps in recent reports~\cite{Sumida_prb2019}
Suppression of Auger scattering that was theoretically predicted to occur in WSMs, could lead to population inversion
 with similar lifetimes~\cite{Afanasiev_prb2019}.

\subsection{Model Hamiltonian with interactions}\label{Hamiltonian}
For the discussion presented in this section, we assume that the lifetime 
of the population inversion, established at $t=t_1$ (see Fig.~\ref{fig1}), is infinitely long so that 
 the system can be considered to be in quasiequilibrium with chemical potentials $\mu_\mathrm{e/h}$ for 
 electron/hole pockets. We will then derive a general Hamiltonian for a pumped DM including Coulomb interactions.

The pumped system in the quasiequilibrium state can be described by the Hamiltonian
\begin{equation}
\mathcal{H}=\mathcal{H}_\mathrm{e}+\mathcal{H}_\mathrm{h}+\mathcal{V},
\label{eq:hamil0}
\end{equation}
where the first two terms represent the non-interacting Hamiltonians for electrons and holes, respectively, and the last term is 
the electron-hole interaction,
written in
the band basis.
The Hamiltonian in Eq.~\ref{eq:hamil0} can be formally derived by diagonalizing the interacting Dirac Hamiltonian $H$ for a 2D or 3D DM and 
by including the finite
chemical
potential for electron and hole bands (the procedure is demonstrated in Ref.~\cite{Pertsova_prb2018} for the case of a DSM/WSM).
Thus the electron and hole Hamiltonians are essentially the linearly-dispersing conduction and valence band of the DM, modified 
by optical pumping
\begin{equation}
\mathcal{H}_{\mathrm{e/h}}=\sum_{\alpha}\sum_{\textbf{k}}\varepsilon^{\mathrm{e/h}}_{\textbf{k}}c^{\alpha\dagger}_{\textbf{k},\mathrm{e/h}}
c^{\alpha}_{\textbf{k},\mathrm{e/h}},
\label{eq:hamil_eh}
\end{equation}
where $\varepsilon^{\mathrm{e/h}}_{\textbf{k}}=\pm\hbar{v}k-\mu_\mathrm{\mathrm{e/h}}$
is the dispersion for the electron/hole populations, $\mu_\mathrm{\mathrm{e/h}}$ is the electron/hole chemical potential, and 
 $v$ is the Fermi velocity of the Dirac states. Here we allow for multiple Dirac cones labeled by index $\alpha$, e.g. $\alpha=1,2$ 
 in graphene or $\alpha=1,2N$, where
 $N$ is an integer, in a WSM;
$c^{\alpha\dagger}_{\textbf{k},\tau}$ ($c^{\alpha}_{\textbf{k},\tau}$) is the creation(annihilation) operator for 
an electronic state with  momentum $\textbf{k}$
 in the
band $\tau=\{\mathrm{e,h}\}$ belonging to the Dirac cone $\alpha$. Equations~(\ref{eq:hamil0})-(\ref{eq:hamil_eh}) 
can be used to study
2D or 3D DMs
by letting $\mathbf{k}$ be a 2D or 3D momentum, respectively.

The original particle-particle interaction term written in the spinor basis is given by
\begin{equation}
 V=\sum_{\sigma,\sigma^{\prime}}\sum_{\mathbf{k},\mathbf{k}',\mathbf{q}}\sum_{\alpha}{\Phi_{\mathbf{k}'+\mathbf{q},\sigma'}^{\alpha_1{\dagger}}}
 \Phi_{\mathbf{k}',\sigma'}^{\alpha_2} {\Phi_{\mathbf{k}-\mathbf{q},\sigma}^{\alpha_3{\dagger}}}
 \Phi_{\mathbf{k},\sigma}^{\alpha_4},
\end{equation}
where $\Phi_{\mathbf{k},\sigma}^{\alpha_i}$ is the spinor corresponding to a state with momentum $\mathbf{k}$ and spin $\sigma$ in a 
Dirac cone $\alpha_i$.
 One obtains the interaction term $\mathcal{V}$ in the band basis by expressing the spinors in the diagonal, or band, basis
 as $\Phi_{\mathbf{q},\sigma}^{\alpha}=\sum_{\tau=e/h}\Phi_{\mathbf{q},\tau,\sigma}^{R/L}$, where
 $\Phi_{\mathbf{q},\tau,\sigma}^{\alpha}=\chi_{\mathbf{q},\tau}^{\alpha,\sigma} c_{\mathbf{q},\tau}^{\alpha}$ and 
 $\chi_{\mathbf{q},\tau}^{\alpha,\sigma}$
 are the eigenvectors of the Dirac Hamiltonian $H$.
The expression for $\mathcal{V}$ depends on the form of $H$ considered and will, in general, inherit the spinor structure 
of the Dirac Hamiltonian.

At the simplest level, considering a spinless model of the DM and only intra-nodal interactions i.e. interactions between
holes and electrons belonging to the same Dirac cone, the interaction term can be written as
\begin{equation}
 \mathcal{V}_0=\sum_{\alpha}\sum_{\textbf{q},\textbf{k},\textbf{k}'} {V}_s({\textbf{q}}) c^{\dagger}_{\textbf{k}+\textbf{q},\mathrm{h}}c^{\dagger}_
 {\textbf{k}^\prime-
 \textbf{q},\mathrm{h}}c_{\textbf{k}',\mathrm{h}}c_{\textbf{k},\mathrm{e}},\label{eq:V0}
\end{equation}
where $V_{s}(\mathbf{q})$ is the screened Coulomb potential. In this case, the number and degeneracy of the Dirac cones will be taken
into account in the form of $V_{s}(\mathbf{q})$. This simplified model was used in Ref.~\cite{Triola_prb2017} to describe
 the transient excitonic insulator phase in 2D DMs.

A more general form of the interaction potential can be derived by considering intra- and inter-nodal excitonic pairing of
the form $\Phi_{\mathbf{q},\mathrm{e},\sigma}^{\alpha\dagger}\Phi_{\mathbf{q},\mathrm{h},\sigma'}^{\beta}$.
For a WSM with broken time reversal symmetry i.e a 3D DM with two
 non-degenerate Dirac nodes (g=2), we adopt the notations  $\mathbf{k}\rightarrow\mathbf{q}$, where  
  $\mathbf{q}=\mathbf{k}\mp\mathbf{K}$, where $\pm\mathbf{K}$ is the position of the two nodes with different chirality. 
The interaction potential for this case in the band basis is given by
\begin{eqnarray}\label{eq:V_WSM}
 \mathcal{V}_W=-\sum_{\mathbf{q},\mathbf{q}'}\sum_{\substack{\tau,\tau\prime=\mathrm{e,h}\\ \tau\ne\tau'}}
 \lbrack\tilde{V}_\mathrm{intra}(\mathbf{q},\mathbf{q}')
 \sum_{\alpha,\beta=1,2}
 c_{\mathbf{q},\tau}^{\alpha\dagger}c_{\mathbf{q},\tau\prime}^{\alpha}
 c_{\mathbf{q}',\tau'}^{\alpha\dagger}c_{\mathbf{q}',\tau}^{\alpha}\\
 +\tilde{V}_\mathrm{inter}(\mathbf{q},\mathbf{q}')
 c_{\mathbf{q},\tau}^{1\dagger}c_{\mathbf{q},\tau'}^{2}c_{\mathbf{q}',\tau'}^{2\dagger}c_{\mathbf{q}',\tau}^{1}\rbrack,
\end{eqnarray}
where $\tilde{\mathcal{V}}_\mathrm{intra}$ and $\tilde{\mathcal{V}}_\mathrm{inter}$ are the intra- and inter-nodal interaction potentials,
respectively,
\begin{eqnarray}
\tilde{V}_\mathrm{intra}(\mathbf{q},\mathbf{q}')&=&V_s(\mathbf{q}-\mathbf{q}')[\frac{\sin\theta\sin\theta'}{2}+\frac{1+\cos\theta\cos\theta'}{2}\nonumber\\
&\phantom{=}&\phantom{{V}_s(\mathbf{q}-\mathbf{q}')[\frac{\sin\theta\sin\theta'}{2}}\times\cos(\phi-\phi')]\label{eq:V_intra}\\
\tilde{V}_\mathrm{inter}(\mathbf{q},\mathbf{q}')&=&-\left[2{V}_s(2\mathbf{K})-{V}_s(\mathbf{q}-\mathbf{q}')(1+\hat{q}\cdot\hat{q}')\right]\label{eq:V_inter},
\end{eqnarray}
Here $\{\hat{q},\theta,\phi\}$ are the polar coordinates and 2$\mathbf{K}$ is the separation between the nodes. 
Note that in Eqs.~\ref{eq:V_intra}-\ref{eq:V_inter}
 we only kept the leading terms in the
 interaction potential by using $|\mathbf{q}-\mathbf{q}'|<<|\mathbf{q}-\mathbf{q}'-2\mathbf{K}|$ and $|\mathbf{q}-\mathbf{q}'|<<|\mathbf{K}|$.

In the next section, we will discuss the form of the screened Coulomb potential ${V}_s(\mathbf{q})$ for a pumped DM.

\subsection{Screened Coulomb interaction}\label{TF_screening}
%
%
%

In the static random phase approximation, the dielectric function $\varepsilon(\mathbf{q},\omega=0)$, 
the screened Coulomb potential $V_s(\mathbf{q})$, 
and the screening wavevector $\kappa$ are given by~\cite{haug2004quantum}
\begin{eqnarray}
  \varepsilon(\mathbf{q},0)&=&\left\{
  \begin{array}{@{}ll@{}}
    1+\frac{\kappa}{q}, & \mathrm{2D} \\
    1+\frac{{\kappa}^2}{q^2}, & \mathrm{3D}
  \end{array}\right.,\label{eq:Lindhart_stat_2D_3D}\\
%
  V_s(\mathbf{q})&=&\left\{
  \begin{array}{@{}ll@{}}
    \frac{V^{\mathrm{2D}}({\mathbf{q}})}{\varepsilon(\mathbf{q},0)}=\frac{2\pi e^2}{\varepsilon}\frac{1}{q+\kappa}, 
    & \mathrm{2D} \\
    \frac{V^{\mathrm{3D}}({\mathbf{q}})}{\varepsilon(\mathbf{q},0)}=\frac{4\pi e^2}{\varepsilon}\frac{1}{q^2+{\kappa}^2}, 
    & \mathrm{3D}
  \end{array}\right.,\label{eq:coul_scr_2D_3D}\\
%
  \kappa&=&\left\{
  \begin{array}{@{}ll@{}}
    \frac{2\pi e^2}{\varepsilon}\sum\limits_{i=\mathrm{e,h}}\frac{\partial n_i}{\partial \mu_i},
    & \mathrm{2D} \\
    \sqrt{\frac{4\pi e^2}{\varepsilon}\sum\limits_{i=\mathrm{e,h}}\frac{\partial n_i}{\partial \mu_i}}, 
    & \mathrm{3D}
  \end{array}\right.,\label{eq:scr_wv_2D_3D}
\end{eqnarray} 
where we have explicitely specified the expressions for a 2D and 3D system. In Eq.~(\ref{eq:coul_scr_2D_3D}), 
 $\varepsilon$ is the dielectric constant of the material and $V^{\mathrm{2D}}({\mathbf{q}})=\frac{2\pi e^2}{\varepsilon}\frac{1}{q}$ 
 ($V^{\mathrm{3D}}({\mathbf{q}})=\frac{4\pi e^2}{\varepsilon}\frac{1}{q^2}$) is the uncreened 
 Coulomb potential in 2D(3D), which is the Fourier transform of the real 
 space Coulomb potential $V(r)=\frac{1}{\varepsilon}\frac{e^2}{r}$.
 
 A system with population inversion consists of electron and hole pockets, which are in general 
 characterized by different densities $n_i$ and chemical potentials $\mu_i$, $i=\mathrm{e,h}$. Hence, in 
 Eq.~(\ref{eq:scr_wv_2D_3D}), we defined the global screening wavevector~\cite{klingshirn2005optics},  
 which combines electron and hole contributions to screening. 
Alternatively, $\kappa$ can be expressed directly in terms of the screening wavevectors for electrons and holes
\begin{eqnarray}
    \kappa&=&\kappa_\mathrm{e}+\kappa_\mathrm{h},\, \kappa_i=\frac{2\pi e^2}{\varepsilon}\frac{\partial n_i}{\partial \mu_i},\quad
     \mathrm{2D} \label{eq:tot_scr_wv_2D}\\
    \kappa^2&=&{\kappa_\mathrm{e}}^2+{\kappa_\mathrm{h}}^2,\, \kappa_i=\sqrt{\frac{4\pi e^2}{\varepsilon}\frac{\partial n_i}{\partial \mu_i}},\quad
     \mathrm{3D}\label{eq:tot_scr_wv_3D}.
\end{eqnarray}

We will now derive the expressions for $\kappa_i$, $i=\mathrm{e,h}$ in the case of DM, i.e. for a system with linear dispersion 
 $E=\hbar{v}k$. We assume a particle-hole symmetric spectrum with identical 
 velocities for electrons and holes, and use the Thomas-Fermi approximation, i.e. $T\rightarrow 0$, in the expression 
 for the screening wavevector~\cite{haug2004quantum,dassarma2011graphene,dassarma2014_3dDirac}. 
 The Fermi wavevector is given by $k^i_{\mathrm{F}}={\mu_i}/\hbar{v}$. 
 In the case of 2D DM, the density is $n_i=N_i/A=\frac{{g}{k^i_{\mathrm{F}}}^2}{4\pi}=\frac{g}{4\pi{v}^2\hbar^{2}}\mu_i^2$, 
 where $N_i$ is the number of quantum states, $A$ is the area, and $g$ is the degeneracy.
In the case of 3D DM, $n_i=N_i/V=\frac{{g}{k^i_{\mathrm{F}}}^3}{6\pi^2}=\frac{g}{6\pi^2{v}^3\hbar^{3}}\mu_i^3$
where $V$ is the system volume. Using the definition of the screening wavevector in 2D and 3D, given in Eqs.~(\ref{eq:tot_scr_wv_2D})-
(\ref{eq:tot_scr_wv_3D}), we 
get 
 \begin{equation}\label{eq:scr_wv_DM}
  \kappa_\mathrm{e/h}=\left\{
  \begin{array}{@{}ll@{}}
    \frac{g e^2}{\varepsilon}\mu_\mathrm{e/h}\equiv{g}\alpha k^\mathrm{e/h}_\mathrm{F},
    & \mathrm{2DDM} \\
    \sqrt{\frac{2 g e^2\varepsilon{v}\hbar}{\pi\varepsilon}}\mu_\mathrm{e/h}\equiv\sqrt{\frac{2{g}\alpha}{\pi}} k^\mathrm{e/h}_\mathrm{F}, 
    & \mathrm{3DDM}
  \end{array}\right..
\end{equation} 

Expressions for electron density, DOS and the screening wavevector 
for 2D and 3D DMs are summarized in Table~\ref{Table:screening}. We also include for comparison 
the results for 2D and 3D electron gas ($E=\hbar^2{k^2}/2m$). For simplicity, we present the 
 results for one type of carriers. The global screening wavevector can be 
 obtained using Eqs.~(\ref{eq:tot_scr_wv_2D})-
(\ref{eq:tot_scr_wv_3D}). In the case of  
 balanced electron and hole populations ($\mu_\mathrm{e}=-\mu_\mathrm{h}$), $\kappa=2\kappa_\mathrm{e/h}$ ($\kappa=\sqrt{2}\kappa_\mathrm{e/h}$) 
 for a 2D(3D) DM.

\begin{table}[ht!]
\caption{Carrier density $n$, density of states $D(E)$, and the screening wavevector $\kappa_\mathrm{TF}$ in 
 the Thomas-Fermi approximation for a DM and a free electron gas (EG) and  
in 2D and 3D. One type of carriers (electrons or holes) is assumed; $m$ is the mass of electron.
}
\begin{tabular}{r||c|c|c}
System  & $n$ &  $D(E)$  & $\kappa_\mathrm{TF}$ \\
\hline
\hline
2D DM & $\frac{g}{4\pi{v}^2\hbar^{2}}\mu^2$  & $\frac{g}{2\pi(\hbar v)^2}E$ & $g\alpha k_\mathrm{F}$\\
\hline
3D DM &  $\frac{g}{6\pi^2{v}^3\hbar^{3}}\mu^3$ &  $\frac{g}{2\pi^2(\hbar v)^3}E^2$ & $\sqrt{\frac{2 g\alpha}{\pi}}k_\mathrm{F}$ \\
\hline
\hline
2DEG  & $\frac{gm}{2\pi\hbar^{2}}\mu^2$ & $\frac{gm}{2\pi\hbar^2}$ & $\frac{{g}e^2{m}}{\varepsilon\hbar^2}$ \\
\hline
3DEG  & $\frac{gm}{6\pi^2\hbar^{3}}(\sqrt{2{m}\mu})^3$ & $\frac{g\sqrt{2m^3}}{{2\pi^2\hbar^3}}E^{1/2}$ & $\sqrt{\frac{2{g}{e^2}{m}}
{\varepsilon\pi\hbar^2}}
k_\mathrm{F}^{1/2}$ \\
\end{tabular}
\label{Table:screening}
\end{table}

As one can see from Table~\ref{Table:screening}, the screnning wavevector in 2D and 3D DMs 
 grows with $\alpha$, $g$ and $k_\mathrm{F}$ or, equivalently, with $\mu$. In both cases, 
 $\kappa$ scales linearly with the chemical potential, however, the prefactors in the linear dependence are different.

\subsection{Gap equation}\label{sec_gap_equation}
\subsubsection{General derivation}\label{sec:Gen_gap_deriv}
Having established the form of the interacting Hamiltonian, we will now proceed with the derivation of the 
  self-consistent equation for the mean-field order parameter, or excitonic gap.  
 In order to obtain the gap equation in the general form, we write down the Hamiltonian of
a pumped DM [Eq.~(\ref{eq:hamil0})], including the interaction term in the band basis
\begin{eqnarray}
\mathcal{H}&=&\sum_{\substack{\mathbf{q} \\ \tau=e,h}}\varepsilon_{\mathbf{q}}^{\tau}c_{\mathbf{q},\tau}^{\dagger}c_{\mathbf{q},\tau}+
\sum_{\substack{\textbf{q},\textbf{q}' \\ \tau\ne\tau'}}
\tilde{V}({\textbf{q}},\textbf{q}') c^{\dagger}_{\textbf{q},\tau}c_{\textbf{q},\tau'}
c^{\dagger}_{\textbf{q}',\tau'}c_{\textbf{q}',\tau}.\label{eq:hamil_band}
\end{eqnarray}
We consider pairing between electrons and holes in 2D or 3D momentum space, with 
 a general interaction potential $\tilde{V}(\mathbf{q},\mathbf{q}')$. For simplicity we have omitted the node subscript $\alpha$.
The interaction term in Eq.~(\ref{eq:hamil_band}) can be easily adjusted to represent
intranodal or internodal interactions for a general system with multiple nodes. More specifically,
 for intranodal interactions, 
 $c_{\mathbf{q},\tau}^{\dagger}\equiv{c}_{\mathbf{q},\tau}^{\alpha\dagger}$ ($c_{\mathbf{q},\tau}\equiv{c}_{\mathbf{q},\tau}^{\alpha}$)
 is the creation(annihilation) operator for the band $\tau=e,h$ of the same node $\alpha$.
 For internodal interactions,
 $c_{\mathbf{q},\mathrm{e}}^{\dagger}\equiv{c}_{\mathbf{q},\mathrm{e}}^{\alpha=1\dagger}$ 
 ($c_{\mathbf{q},\mathrm{e}}\equiv{c}_{\mathbf{q},\mathrm{e}}^{\alpha=1}$) and
 $c_{\mathbf{q},\mathrm{h}}^{\dagger}\equiv{c}_{\mathbf{q},\mathrm{h}}^{\alpha=2\dagger}$ 
 ($c_{\mathbf{q},\mathrm{h}}\equiv{c}_{\mathbf{q},\mathrm{h}}^{\alpha=2}$)
 are, respectively, the creation(annihilation) operators for the conduction and valence bands belonging to two different nodes; 
 also note that momentum $\mathbf{q}$ refers to a specific node, $\mathbf{q}\equiv\mathbf{k}\mp\mathbf{K}$,where $\pm\mathbf{K}$ is the position 
 of the node. 
 The interaction potential $\tilde{V}(\mathbf{q},\mathbf{q}')$ takes on the form derived in Section~\ref{Hamiltonian}
 for intra- or internodal pairing, Eqs.~(\ref{eq:V_intra})-(\ref{eq:V_inter}).

Furthermore, we introduce the electron(hole) Green's functions and the anomalous Green's function 
\begin{eqnarray}
G_{\mathrm{e(h)}}(\mathbf{q},t-t') & = & -<T_{t} c_{\mathbf{q},\mathrm{e(h)}}(t)c^{\dagger}_{\mathbf{q},\mathrm{e(h)}}(t)>,\\
F(\mathbf{q},t-t') & = & -<T_{t} c_{\mathbf{q},\mathrm{e}}(t){c_{\mathbf{q},\mathrm{h}}}^{\dagger}(t')>,
\end{eqnarray}
where $T_{t}$ is the imaginary time-ordering operator. In the mean-field approximation, the order parameter, or 
excitonic gap $\Delta(\mathbf{q})$, is calculated as
\begin{eqnarray}\label{eq:gap_def}
 \Delta({\textbf{q}})&=&\sum_{\textbf{q}'}\tilde{V}({\textbf{q},\textbf{q}'})\langle c_{\textbf{q}',\tau}c^\dagger_{\textbf{q}',\tau'}
 \rangle\nonumber\\
 &=&T\sum_{\textbf{q}',i\omega_n}\tilde{V}({\textbf{q},\textbf{q}'})F(\textbf{q}';i\omega_n),
\end{eqnarray}
where $\omega_n=(2n+1)\pi/\beta$ is a fermionic Matsubara frequency, $\beta=1/k_{\mathrm{B}}T$, $k_\mathrm{B}$
 is the Boltzmann constant and $T$ is the temperature assumed to be
the same for photoexcited electrons and holes. 
In order to obtain the gap equation, one can 
 use, for example, the Gor'kov approach,
which is based on time-dependent equations of motions for $G_\mathrm{e(h)}$
and $F$~\cite{excitons}. The final result for the gap equation reads
\begin{equation}
 \Delta(\mathbf{q})=\sum_{\mathbf{q}'}\tilde{V}(\mathbf{q},\mathbf{q}')\frac{\Delta(\mathbf{q}')}{\omega_{+}({\mathbf{q}'})-
 \omega_{-}({\mathbf{q}'})}
 [n_{\mathrm{F}}(\omega_{+})-n_{\mathrm{F}}(\omega_{-})],
\end{equation}
where
\begin{equation}
 \omega_{\pm}(\mathbf{q})=\frac{\varepsilon^\mathrm{e}_{\mathbf{q}}+\varepsilon^\mathrm{h}_{\mathbf{q}}}{2}\pm\frac{1}{2}
 \sqrt{(\varepsilon^\mathrm{e}_{\mathbf{q}}-\varepsilon^\mathrm{h}_{\mathbf{q}})^2+4|\Delta(\mathbf{q})|^2}
\end{equation}
are the renormalized excitonic bands and $n_{\mathrm{F}}(\omega)=1/(e^{\beta\omega}+1)$ is the Fermi-Dirac distribution. 
In equilibrium, i.e. in the limit $\mu_\mathrm{e}=-\mu_\mathrm{h}=0$,  
one recovers the usual BCS gap equation.

\subsubsection{Intranodal vs interdonal interactions}\label{sec:EX_CDW_gap}
%
%
Using the results of the previous section, we will discuss the
competing excitonic phases originating from internodal and
intranodal interactions for the specific case of 3D DMs. 
Note that here we consider either a DSM or a WSM with broken time reversal symmetry.  
Taking into account different interations present in the system,
we can specify the form of the general order parameter introduced in Eq.~(\ref{eq:gap_def})
\begin{equation}\label{eq:gap_def_EI_CDW}
 \Delta(\mathbf{q})=\sum_{\mathbf{q}'}\tilde{V}(\mathbf{q},\mathbf{q}')\left\langle
 c^{\alpha\dagger}_{\mathbf{q}',\tau} c^{\beta}_{\mathbf{q}',\tau'} \right\rangle.
\end{equation}
For $\alpha=\beta$, $\Delta(\mathbf{q})$ describes pairing between between electrons and 
holes in the same node (intranodal interactions), while for $\alpha\ne \beta$, pairing occurs  
between electrons and holes belonging to different nodes (internodal interactions). 
 Intranodal pairing leads to an excitonic insulator (EI) phase. Internodal pairing gives rise to  
a charge density wave (CDW) phase with the modulation momentum equal to the distance $2\mathbf{K}$
between the nodes. Equilibrium excitonic phases in a WSM with broken time reversal symmetry were studied 
in Refs.~\cite{Wei_prl2012,Wei_prb2014}. In the case of short-range (contact) interaction potential, 
the EI phase is more energetically favorable. In the case of unscreened Coulomb potential, 
the CDW phase becomes dominant. We will show in Section~\ref{sec:tunability} that this result holds in the 
case of the screened Coulomb potential in a pumped WSM, with an important difference that 
both the EI phase and the CDW phase are present at arbitrary weak coupling.

We will further specify the form of the order
parameter and the gap equation for intranodal and internodal interactions in a pumped system. 
The intranodal part of the interaction potential was introduced in Eq.~(\ref{eq:V_intra}).
Keeping only the slowest varying term in the angular-dependent part of the potential proportional to $\cos(\phi-\phi')/2$,
we get $\tilde{V}_\mathrm{intra}=V^s(\mathbf{q}-\mathbf{q}')\cos(\phi-\phi')/2$.
The self-consistent gap equation becomes
\begin{eqnarray}
 \Delta^{\alpha}(\mathbf{q})&=&
\sum_{\mathbf{q}'}2\tilde{V}_\mathrm{intra}\frac{\Delta(\mathbf{q}')}{\omega_{+}({\mathbf{q}'})-\omega_{-}({\mathbf{q}'})}
 [n_{\mathrm{F}}(\omega_{+})-n_{\mathrm{F}}(\omega_{-})],\nonumber\\
 &=&\frac{1}{(2\pi)^3}\int{V}^s(\mathbf{q}-\mathbf{q}')\frac{\Delta^{\alpha}(\mathbf{q}')\cos(\phi-\phi')}
  {\omega_{+}({\mathbf{q}'})-\omega_{-}({\mathbf{q}'})}\nonumber\\
  &\phantom{=}&\phantom{\frac{1}{(2\pi)^3}\int{V}^s(\mathbf{q}-\mathbf{q}')}\times
  [n_{\mathrm{F}}(\omega_{+})-n_{\mathrm{F}}(\omega_{-})]dV,\label{eq:EI_gap}
\end{eqnarray}
and the excitonic gap is given by
\begin{equation}
 \Delta^{\alpha}(\mathbf{q})=\sum_{\mathbf{q}'}\tilde{V}_\mathrm{intra}
 (\mathbf{q},\mathbf{q}')\left\langle c_{\mathbf{q}',\mathrm{e}}^{\alpha\dagger}c_{\mathbf{q},\mathrm{h}}^{\alpha}\right\rangle,\quad\alpha=1,2.
 \label{eq:gap_intra_definition}
 \end{equation}
The mean-field Hamiltonian of the system with intranodal interactions reads
\begin{eqnarray}
 H_\mathrm{intra}&=&\sum_{\mathbf{q},\alpha,\tau}
 \varepsilon_{\mathbf{q}}^{\tau}c_{\mathbf{q},\tau}^{\alpha\dagger}c_{\mathbf{q},\tau}^{\alpha}
 -\sum_{\mathbf{q},\alpha}\tilde{\Delta}^{\alpha}(\mathbf{q})c_{\mathbf{q},\mathrm{h}}^{\alpha\dagger}c_{\mathbf{q},\mathrm{e}}^{\alpha}\nonumber\\
 &-&\sum_{\mathbf{q},\alpha}\tilde{\Delta}^{\alpha*}(\mathbf{q})c_{\mathbf{q},\mathrm{e}}^{\alpha\dagger}c_{\mathbf{q},\mathrm{h}}^{\alpha},\label{eq:EI_Hamil}
\end{eqnarray}
where $\tilde{\Delta}^{\alpha}(\mathbf{q})=2\Delta^{\alpha}(\mathbf{q})$. The first term in Eq.~(\ref{eq:EI_Hamil}) is the non-interacting
Hamiltonian of the two nodes while the last two terms describe interactions within each node.

%
The internodal part of the interaction potential is given by Eq.~(\ref{eq:V_inter}). Furthermore, by 
neglecting the angular-dependent part proportional to $\hat{q}\cdot\hat{q}'$, the interaction potential 
is simply the screened Coulolb potential $V_s(\mathbf{q}-\mathbf{q}')$. In this case, the self-consistent gap equation reads
\begin{eqnarray}
 \Delta^{\tau}(\mathbf{q})&=&\sum_{\mathbf{q}'}\tilde{V}_\mathrm{inter}(\mathbf{q},\mathbf{q}')
 \frac{\Delta(\mathbf{q}')}{\omega_{+}({\mathbf{q}'})-\omega_{-}({\mathbf{q}'})},\nonumber\\
 &=&\frac{1}{(2\pi)^3}\int{V}^s(\mathbf{q}-\mathbf{q}')\frac{\Delta_n(\mathbf{q}')}{\omega_{+}({\mathbf{q}'})-\omega_{-}({\mathbf{q}'})}\nonumber\\
 &\phantom{=}&\phantom{\frac{1}{(2\pi)^3}\int{V}^s(\mathbf{q}-\mathbf{q}')}
 \times[n_{\mathrm{F}}(\omega_{+})-n_{\mathrm{F}}(\omega_{-})]dV,\label{eq:CDW_gap}
\end{eqnarray}
and the excitonic gap is given by
\begin{equation}
 \Delta^{\tau}(\mathbf{q})=\sum_{\mathbf{q}'}\tilde{V}_\mathrm{inter}(\mathbf{q},\mathbf{q}')
 \left\langle c_{\mathbf{q}',\tau'}^{R\dagger}c_{\mathbf{q},\tau}^{L}\right\rangle,\quad\tau=\mathrm{e,h},\quad\tau'\ne\tau.
\end{equation}
The corresponding mean-field Hamiltonian reads
\begin{eqnarray}
 H_\mathrm{inter}&=&\sum_{\mathbf{q},\alpha,\tau}\varepsilon_{\mathbf{q}}^{\tau}c_{\mathbf{q},\tau}^{\alpha\dagger}
 c_{\mathbf{q},\tau}^{\alpha}-
 \sum_{\mathbf{q},\tau\ne\tau'}\Delta_{\tau}(\mathbf{q})c_{\mathbf{q},\tau}^{L\dagger}c_{\mathbf{q},\tau'}^{R}\nonumber\\
 &-&\sum_{\mathbf{q},\tau\ne\tau'}\Delta_{\tau}^{*}(\mathbf{q})c_{\mathbf{q},\tau'}^{R\dagger}c_{\mathbf{q},\tau}^{L}.\label{eq:CDW_Hamil}
\end{eqnarray}
As in Eq.~\ref{eq:EI_Hamil}, the first term in Eq.~(\ref{eq:CDW_Hamil}) is the non-interacting 
part of the Hamiltonian, while the last two terms represent interactions 
between the nodes. 

The CDW gap equation, Eq.~(\ref{eq:CDW_gap}), does not contain any explicit angular dependence (apart 
from the screened Coulomb potential $V_s(\mathbf{q}-\mathbf{q}')$, which depends on $\mathbf{q}-\mathbf{q}'$); hence the corresponding order
parameter is isotropic. In contrast to this, the EI gap equation, Eq.~(\ref{eq:EI_gap}), contains 
a factor $\cos(\phi-\phi')$. By taking $\cos(\phi-\phi')=(e^{i(\phi-\phi')}+e^{i(\phi'-\phi)})/2$ and 
 re-defining the EI gap as $\Delta^{\alpha}(\mathbf{q})=\Delta^{\alpha}(q)\cdot e^{i\phi}$, 
 one obtains a self-consistent 
equation for the magnitude of the EI gap  $\Delta^{\alpha}(q)$, which is identical to the CDW gap equation with an additional
factor of $1/2$. Hence, the EI gap is always smaller than the
 CDW gap for the same model parameters. (One exception is the special (hypothetical) case $g=1$, for which 
 only one node has population inversion and therefore only intranodal interactions leading to the EI phase are relevant). 
 Furthermore, the screened Coulomb potential in the gap equation can be replaced by its angle averge,  
 ${V}_s({\mathbf{q}-\mathbf{q}'})\rightarrow\left\langle{V}_s\right\rangle_{\phi,\theta}(q,q')$, which   
  depends only on the magnitudes of vectors $\mathbf{q}$ and $\mathbf{q}'$.

%

Since the order parameter is momentum-dependent, we will solve the self-consistent gap equation 
 numerically on a finite momentum mesh. We will use the dimensionless units $\Delta\rightarrow\Delta/\hbar{v}\lambda$,
  $\mathbf{q}\rightarrow\mathbf{q}/\lambda$,
  $T\rightarrow{k_\mathrm{B}T}/{\hbar{v}\lambda}$, where $\lambda$ is the momentum cutoff of the Dirac Hamiltonian in Eq.~\ref{eq:hamil_eh}.
The corresponding cutoff energy scale is $\Lambda\equiv\hbar{v}\lambda$. 
 From the solutions of the gap equation at zero and finite temperatures, we will 
 construct the phase diagrams of the trasient excitonic condensate and calculate the observable 
 quantities such as the spectral function and DOS.

\section{Properties of transient excitonic condensate}\label{exp:sig}
%
Assuming a total number of nodes $N$ in the system, one can consider the following cases for optical pumping:
(i) \textit{uniform pumping}, i.e. all nodes are pumped and population 
inversion is realized for each node, 
and (ii) \textit{selective pumping}, i.e. population inversion is realized for a certain number of
nodes $N_0\le{N}$. For $N>1$, pairing within a single node is affected by screening of
carriers belonging to other nodes, provided there is a finite density of electrons and holes at the node. 
In the expression for the Thomas-Fermi screening wavevector [see Table~\ref{Table:screening}],  the 
degeneracy factor $g$ essentially represents the number of cones which contribute to screening. Importantly, 
screening becomes stronger for larger $g$.

For a Bi$_2$Se$_3$-type 3D TI with a single Dirac node on the surface ($N=1$), we have $g=1$. 
For a WSM with two non-degenerate nodes ($N=2$), case (i) corresponds to $g=2$, while case (ii) corresponds to $g=1$ 
in the calculation of teh screening wavevector. In graphene,
 there is an additional spin degeneracy of the Dirac states and therefore
 case (i) corresponds to $g=4$, while case (ii) corresponds to $g=2$ (valley-selective pumping). 
3D DMs can have a large number of nodes, for example $N=24$ in TaAs WSM. In such systems, if all nodes are uniformly pumped,
excitonic effects will be inhibited by screening.

Furthermore, for systems with $N>1$, it is important to clarify 
 which type of interactions is relevant 
in the two pumping cases. For the simplest case of a WSM with two nodes and for uniform pumping ($g=2$),
 both intranodal and internodal interactions are present and therefore both the EI and CDW phases are realized. 
 For selective pumping on a single Weyl node 
 ($g=1$),  one of the nodes has a population inversion, while the other one is in equilibrium. Throughout this work, 
 we assume that the equilibrium chemical 
 potential is located at the node, leading to vanishing DOS. 
 In this case, the strongest pairing is realized for intranodal interactions within the pumped node, 
 leading to the EI phase. 
 The CDW gap rapidly vanishes as a function of
the mismatch between the equilibrium and non-equilibrium chemical potentials. 
Similarly, the EI gap vanishes for $\mu_\mathrm{e}\ne-\mu_\mathrm{h}$~\cite{Triola_prb2017}. 
For a DSM with the minimal degeneracy $g=2$,  only intranodal
interactions are included. For a DSM with $g\ge 4$, both intranodal and internodal
interactions are relevant.

\subsection{Spectroscopic signatures}
\begin{figure*}[ht!]
\includegraphics[width=0.98\linewidth,clip=true]{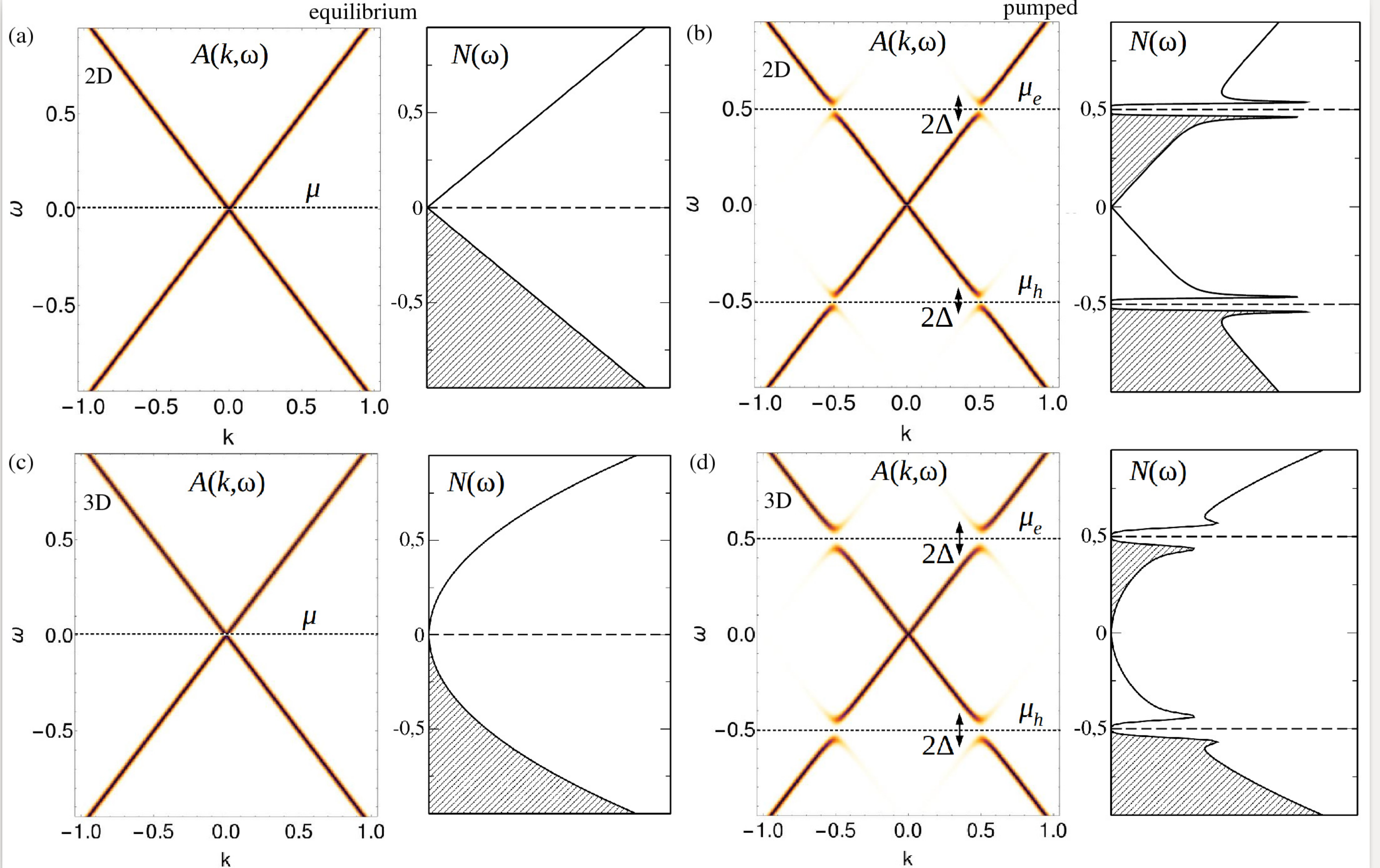}
\caption{The calculated spectral function $A(k,\omega)$ and DOS $N(\omega)$, as functions of 
momentum and energy for a 2D (a,b) and 3D DM (c,d) in the normal state and
in the transient excitonic insulator phase. For illustrative purposes, in the case of 2D DM we use
$g=1$ and $\alpha=1$ which corresponds to
valley-selective pumping in graphene on the substrate; in the case of 3D DM, we use $g=1$ and $\alpha=3$ which
corresponds to pumping on a single Dirac cone in a WSM in the EI phase.
In both 2D and 3D cases, $\mu_\mathrm{e}=-\mu_\mathrm{h}=0.5$ and
the cutoff energy scale is $\Lambda=1$~eV.
Energy and momentum are in dimensionless units.}
\label{fig2}
\end{figure*}

In the pumped DM, the excitonic phase is characterized by energy gaps that open up at
the non-equilibrium chemical potentials (Fig.~\ref{fig2}). 
In the DOS and the spectral function, the gaps separate filled states below 
 from the empty states above the  chemical potentials.  
 The vanishing DOS in the gaps could be detected by scanning tunneling microscopy. 
 The spectral function can be probed by trARPES. 
 A feature that could be directly detected in the ARPES spectra is the bending
 and vanishing of the spectral weight of the 
occupied bands near the non-equilibrium chemical potentials.

The excitonic gap in a pumped DM is a dynamical quantity 
 due to the transient nature of the photoexcited Dirac states. As the
 electron and hole chemical potentials approach the equilibrium value ($\mu=0$ before pumping)
 over the lifetime of  the population inversion, the position of the
 gap will move towards the node and the size of the gap will decrease. This is shown schematically
in Fig.~\ref{fig1}. A more detailed discussion of
the timescales involved and the relaxation of the order parameter
based on rate equation is presented in Section~\ref{sec:dynamics}.
\subsection{Tunability of the critical coupling}\label{sec:tunability}
In equilibrium DM, there is a critical value $\alpha_c^0$ of the dimensionless coupling constant such that for
 $\alpha>\alpha_c^0$ the system becomes an excitonic insulator~\cite{kotov2012electron}. 
 In this section, we will discuss the 
 reduction of the critical coupling which is expected in pumped DMs due to the enhanced DOS away from the Dirac node.

In the case of 2D DM, analytical analysis of the gap equation has been carried out
 in the limit of strong screening, when 
 the Coulomb potential becomes a contact interaction in position space or, equivalently, a constant in momentum space 
$V^s(\textbf{q})\xrightarrow[\kappa>>\lambda]{ }V_0=2\pi v \alpha/\kappa$~\cite{Triola_prb2017}, where $\kappa$ is the screening wavevector.
The important result of the 
analytical model is that for perfectly matched chemical potentials ($\mu_\mathrm{e}=-\mu_\mathrm{h}$),
 the exitonic insulator phase is realized for any finite value of $\mu_\mathrm{e/h}$. The
 calculated excitonic gap $\Delta$ and the critical temperature $T_c$ increase monotonically with increasing
 $\mu_\mathrm{e/h}$. 

 In the case of the realistic screened Coulomb potential, 
 the phase diagram for the order parameter is more complex due to the 
 competition between screening and the enhanced DOS for a given 
$\mu_\mathrm{e/h}$. 
As a result of this competition, 
pumping is more advantageous for excitonic condensation compared to equilibrium only in a certain region of the parameter 
space defined by material parameters, $\alpha$, $\mu_\mathrm{e/h}$ and $g$. This crucial point is demonstrated below for
 2D and 3D DMs.
 
We consider pumping to be efficient if (i) for $\alpha<\alpha_c^0$, the transient excitonic gap is different from zero, or 
(ii) for $\alpha>\alpha_c^0$, the transient excitonic gap is larger than the equilibrim gap. 
In what follows we introduce for convenience a 
single quasi-equilibrium chemical potential $\bar{\mu}=\mu_\mathrm{e}=-\mu_\mathrm{h}$.
 Population inversion with balanced chemical potentials can be realized if 
 the  chemical potential is positioned at the Dirac point before pumping.

\begin{figure}[ht!]
\centering
\includegraphics[width=0.98\linewidth,clip=true]{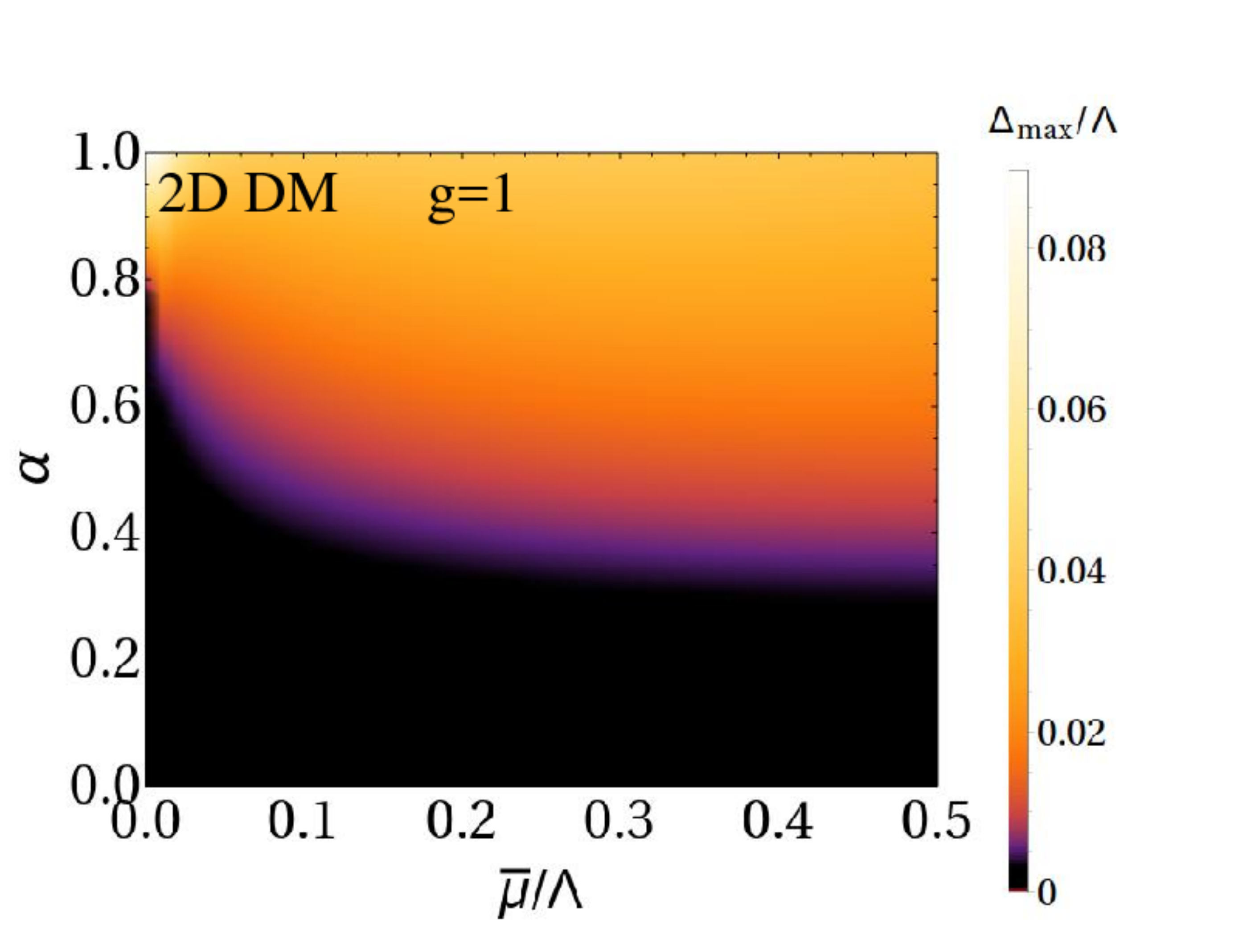}
\caption{Tunability of the critical coupling in 2D DM: 
the maximum value of the excitonic gap as a function 
of $\bar{\mu}$ and $\alpha$ for a 2D DM with $g=1$ and for $T=0$. 
 The color shade
 represents the absolute value of the gap and is logarithmically scaled for better contrast.
Energy is in dimensionless units.}
\label{fig3}
\end{figure}

\begin{figure}[ht!]
\centering
\includegraphics[width=0.98\linewidth,clip=true]{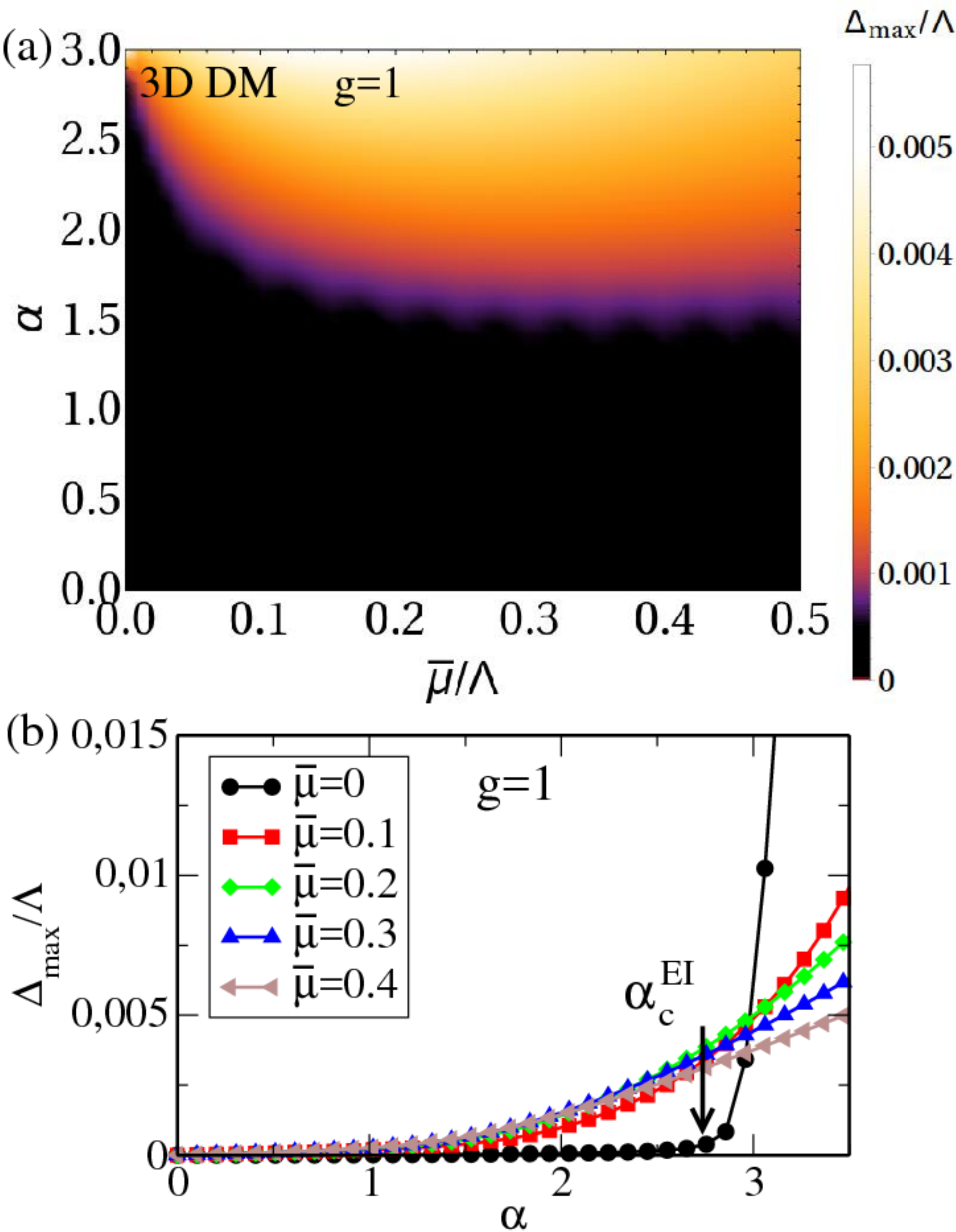}
\caption{Tunability of the critical coupling in 3D DM: 
(a) the maximum value of the excitonic gap as a function
of $\bar{\mu}$ and $\alpha$ for a WSM with $g=1$ (selective pumping)
for $0\le\alpha\le{3}$ and for $T=0$; (b) the maximum value of the gap
as a function of $\alpha$ for the same parameters as in panel (a) and for different values of $\bar{\mu}$. 
 In a WSM with $g=1$, the EI phase dominates (see text). The color shade
in panel (a) represents the absolute value of the gap and is logarithmically scaled for better contrast.
Energy is in dimensionless units.}
\label{fig4a}
\end{figure}

Figure~\ref{fig3} shows the phase diagram of the transient excitonic state in the $\bar{\mu}-\alpha$ plane for a 2D DM. 
Here we consider the regime $\alpha\lesssim{\alpha_c^{\mathrm{2D}}}$, where $\alpha_c^{\mathrm{2D}}\approx{1.0}$ is
 the equilibrium critical coupling in 2D. In equilibrium, the Coulomb potential becomes unscreened when $\mu=0$. In this case, for numerical 
 calculations with the Thomas-Fermi model we use a small but finite value of the screening wavevector such 
 that variations around this value do not change appreciably the result for $\alpha_c^{0}$.  
 The critical coupling in a pumped system is defined for a given $\bar{\mu}$ as the minimum value of $\alpha$ for 
which the gap is different from zero. In practice, we 
 use a condition $\Delta_\mathrm{max}\ge\delta$, where $\Delta_\mathrm{max}$ is the maximum value of the gap and 
 $\delta$ is a small number ($\delta=10^{-6}$ in units of energy). 
 In Figs.~\ref{fig3} and ~\ref{fig4a}(a), the critical coupling defines a line separating the dark ($\Delta_\mathrm{max}=0$)
 and bright ($\Delta_\mathrm{max}\neq{0}$) regions of the phase diagram.  
 As one can see from Fig.~\ref{fig3}, the critical coupling is significantly reduced 
 in the pumped state, where $\bar{\mu}\neq 0$. 
 A similar behavior is found for a 3D DM, as demonstrated in Fig.~\ref{fig4a}(a) for a WSM with $g=1$ and
 for $\alpha$ in the range $0\le\alpha\le{3}$.

\begin{figure}[ht!]
\centering
\includegraphics[width=0.98\linewidth,clip=true]{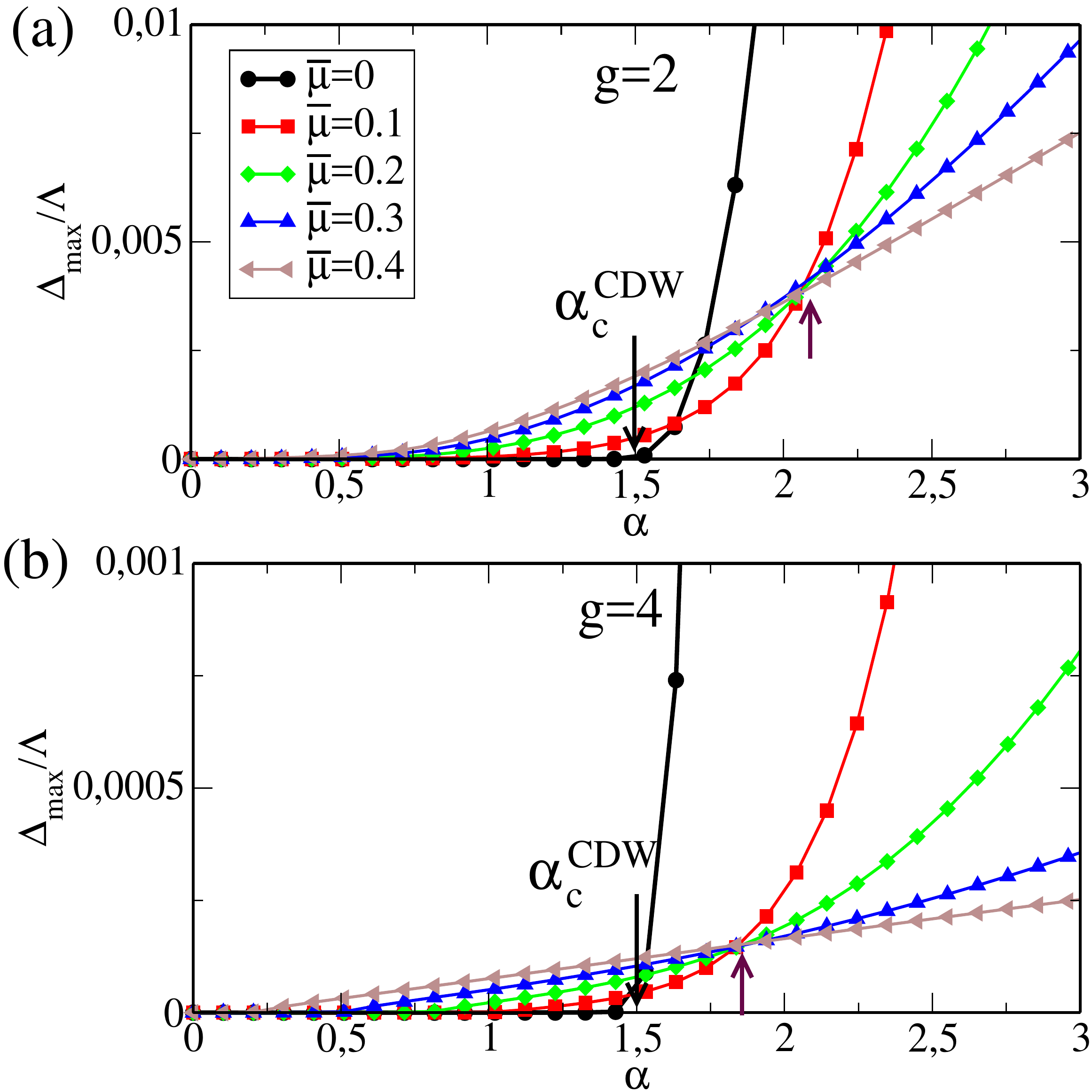}
\caption{The maximum value of the excitonic gap 
as a function of $\alpha$ for different values of $\bar{\mu}$ for a WSM with (a) $g=2$ and (b) $g=4$ and for $T=0$. 
 In a WSM with $g>1$, the CDW phase dominates. Energy is in dimensionless units.}
\label{fig4b}
\end{figure}

The tunability of the critical coupling with pumping is further confirmed in Fig.~\ref{fig4a}(b) and Fig.~\ref{fig4b}
which show the scans of the $\alpha-\bar{\mu}$ phase diagram ($\Delta_\mathrm{max}(\alpha)$ curves),
 for a 3D DM with $g=1$, $2$ and $4$ for increasing values of the chemical potential $\bar{\mu}$.
In equilibrium ($\bar{\mu}=0$), the values of the critical coupling for the
CDW and the EI phases are given by $\alpha_c^{\mathrm{EI}}\approx{3.0}$ and $\alpha_c^{\mathrm{CDW}}\approx{1.5}$, respectively, 
 in agreement with the values obtained analytically in Ref.~\cite{Wei_prb2014}.

 In the pumped case ($\bar{\mu}\ne{0}$), the critical coupling for excitonic instability 
vanishes. In the range $\alpha\lesssim{\alpha_c^{\mathrm{CDW(EI)}}}$,  
 the equilibrium system remains gapless, while the pumped system develops a gap. 
 This is the regime where optical pumping is efficient, regardless of the 
 values of other material parameters. 
As one can see from  Fig.~\ref{fig4a}(b) and Fig.~\ref{fig4b}(a,b), there is also a narrow range of $\alpha$ just above the equilibrium 
critical coupling, where one can find a value of $\bar{\mu}$ for which the transient excitonic gap is larger than the 
equilibrium gap. This range shrinks with increasing the generacy factor $g$, due to stronger screening. 
Since the size of the gap in equilibrium is controlled only by $\alpha$, 
the equilibrium excitonic gap grows rapidly with increasing $\alpha$ and becomes larger than the transient excitonic gap. 

Moreover, we find that in the range $\alpha\lesssim{2.0}$ and for $\bar{\mu}\neq{0}$, 
the size of the gap increases monotonically with increasing $\bar{\mu}$.  
For larger $\alpha$ the dependence of the gap on $\bar{\mu}$ becomes non-monotonic. 
 For $\alpha\approx{2}$, which is marked by upwards pointing arrows in Fig.~\ref{fig4b},
  the size of gap for a given $\alpha$ starts to decrease with increasing the chemical potential.
  This indicates a point where screening effects win over the enhancement of DOS at the non-equilibrium chemical potentials,
  making pumping inefficient. Figure~\ref{fig4b} also shows that the size of the gap
decreases with increasing $g$.

\subsection{Phase diagrams of the excitonic condensate}
\begin{figure*}[ht!]
\centering
\includegraphics[width=0.98\linewidth,clip=true]{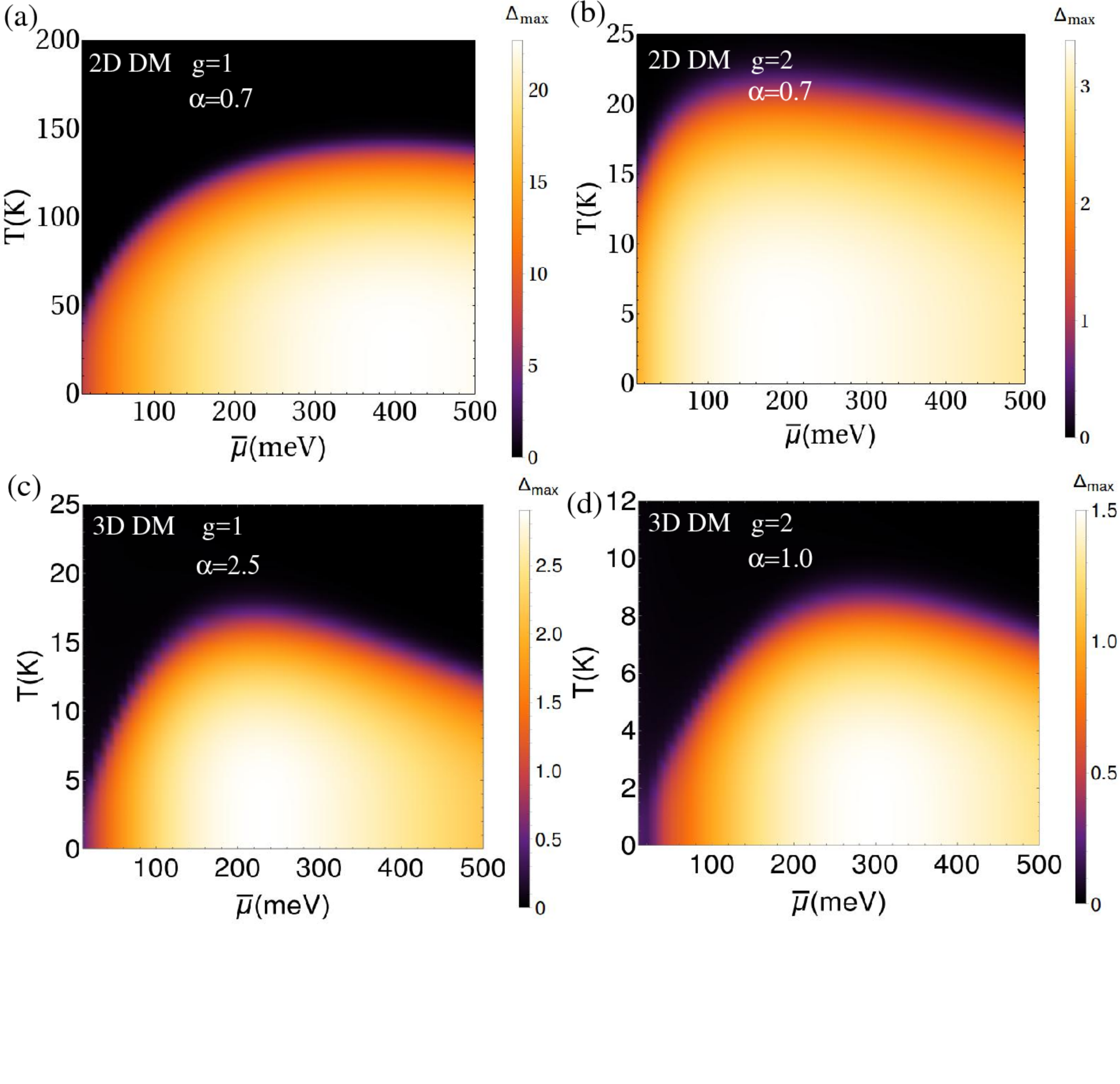}
\caption{
The maximum value of the excitonic gap as a function 
of $\bar{\mu}$ and $T$. 
Top row is for a 2D DM with $\alpha=0.7$ for (a) $g=1$ and (b) $g=2$. 
Bottom row is for a 3D DM (WSM) for (c) $g=1$ and $\alpha=2.5$ and (d) $g=2$ and $\alpha=1$. Note
that for WSM with $g=1$, only the EI phase is realized. For WSM with $g=2$ both EI and CDW phases are present,
however the CDW gap is the largest. Hence, $\alpha$ is chosen to be smaller than the corresponding equilibrium
critical coupling in panels (c) and (d). The color shade represents the absolute value of the gap in meV's.
The energy cutoff is $\Lambda=1$~eV in both 2D and 3D cases. Note that here we retain physical units for
energy and temperature to make a connection with estimates for $\Delta_\mathrm{max}$ and $T_c$ in Table~\ref{tab:values}.}
\label{fig5}
\end{figure*}

Figure~\ref{fig5} illustrates the effect of pumping on the $T-\bar{\mu}$ phase diagrams. 
The critical temperature $T_c$ is defined 
as the minimum value of $T$ for which  
 the excitonic gap is different from zero. 
In Fig.~\ref{fig5}, the following cases are considered: (i) a 2D DM with two nodes and $\Lambda=1$~eV, which corresponds
to graphene on the substrate (top panels), and (ii) a hypothetical WSM with two nodes and $\Lambda=1$~eV (bottom panels).
We investigate selective ($g=1$) and uniform ($g=2$) pumping.
In both cases, the values of $\alpha$ are chosen to be smaller than the equilibrium critical
coupling: $\alpha<\alpha_c^{2D}\approx{1}$ for graphene and $\alpha<\alpha_c^{\mathrm{CDW/EI}}$ for WSM.
According to Figs.~\ref{fig2}-\ref{fig4b}, this is the 
coupling regime in which pumping promotes the excitonic instability.

In Fig.~\ref{fig5}, the line separating the dark 
($\Delta_\mathrm{max}=0$) and bright ($\Delta_\mathrm{max}\neq{0}$) regions of the phase diagram defines the dependence of $T_c$ on 
$\bar{\mu}$. The behavior of $T_c$ is similar in all cases considered:
$T_c$ increases monotonically with the chemical potential until it reaches a maximum at a certain value of $\bar{\mu}$;
further increasing 
$\bar{\mu}$ leads to a deacrease in $T_c$ due to screening. The peak in $T_c$ as a function of $\bar{\mu}$ determines
the regime in which optical pumping is most efficient. The value of the chemical potential that gives the largest $T_c$
 is determined by the material specific constants $\alpha$ and $g$.
 For exampled, for a 2D DM and the same value of $\alpha$, the peak shifts to smaller values of $\bar{\mu}$ due to larger screening.
 For a WSM, the downturn in $T_\mathrm{c}$ for $g=2$ occurs
 at a slightly larger $\bar{\mu}$ compared to $g=1$ due to different values of $\alpha$ considered in the two cases.

\section{Discussion and Experimental feasibility}\label{exp:mater}
\subsection{Comparison to equilibrium DMs with population inversion}
Before discussing experimental feasibility of excitonic condensate in pumped DM,
 we should note that finite density of electrons and holes
can be realized in equilibrium systems. 
In the 2D case, parallel
 magnetic field applied to a single layer of graphene creates two perfectly nested electron and hole Fermi surfaces,
 allowing for condensation~\cite{aleiner_prb2007}. Finite equilibrium populations can be achieved in double layer systems, such
 as double layer graphene~\cite{Zhang_prb2008,Min_prb2008,Kharitonov_prb2008} and TI thin films~\cite{Seradjeh_prl2009,
 Tilahun_prl2011}, where the chemical potentials
  of the two ``layers'' are tuned independently by using electrostatic gates.

  A similar situation
   can be realized in materials with multiple Dirac/Weyl nodes, where the energies of at least two nodes
  are not required to be the same. A specific example in 2D is a
  topological Kondo insulator SmB$_6$ at finite chemical potential, provided there is a sufficient energy
  offset between the nodes at $\Gamma$ and $X$/$Y$ points~\cite{Roy_prb2015}.  In 3D,
perfectly nested electron and hole Fermi surfaces can be achieved in  WSM with broken spatial inversion
symmetry~\cite{zyuzin_prb2012_ex_wsm}.

An important difference from earlier work on double layer 2D systems is that our theory of pumped DM applies
to a single layer DMs, i.e. monolayer graphene or TI surface. In this case, finite populations of electrons
 and holes cannot be made by gating alone. Furthermore,
 as will be shown below, the effect of metallic screening can be minimized in single layer DMs.
  In double layer graphene,
 the number of fermions involved with screening is twice that of single layer graphene and therefore the maximum value
 of the coupling constant decreases by a factor of 2, this sets an upper limit on the critical temperature of
 $1mK$~\cite{Kharitonov_prb2008}.
 For single layer DMs the maximum value of the coupling constant increases by at least a factor of
 2 (4, in the absence of the valley degree of freedom, as in 3D TIs). This leads to an exponentially
 enhanced upper critical temperature relative to the double layer case. Furthermore, since the system is a single layer,
 the interaction is not exponentially suppressed with increasing the distance between layers as in the case of the double
 layer system.

More generally, optical pumping is a novel way of generating excitonic states that have a transient nature. In the case of DMs,
it also offers a way of enhancing and effectively controlling the strength of interaction between
non-equilibrium electron and hole populations.

\subsection{Estimates for excitonic gaps and $T_c$ in pumped DMs}
Numerical  estimates
of the critical temperature and the size of the excitonic for several examples of realistic and hypothetical 2D and 3D DMs
 are summarized in Table~\ref{tab:values}.

\begin{table*}[ht]
\caption{Estimates of the critical temperature $T_c$ and the maximum of the excitonic gap $\Delta_\mathrm{max}$
in pumped 2D and 3D DMs. Material-specific parameters, i.e. the dimensionless coupling constant $\alpha$, 
the cutoff energy scale $\Lambda$  and the lifetime of the transient excitonic
state $\tau$ were estimated based on existing literature. 
For graphene and 3D TIs, the average chemical potential $\bar{\mu}$ used in the numerical calculation of the gap was
estimated based on experimental data. For all other cases, the values of $\Delta_\mathrm{max}$ and $T_c$ were 
obtained by taking $\bar{\mu}=\Lambda/2$. The last row in the section of the table
corresponding to 2D DMs shows the estimates for a hypothetical 2D DM with $g=1$ and with other parameters similar to graphene.
For hypothetical 3D DMs (last three rows of the table) the cutoff energy scale $\Lambda=1$~eV is assumed.} 
\label{tab1}
\begin{tabular}{r||l|l|l|c|r}
Material &  $\alpha$ &  $\Lambda$ (eV) & $\tau$ (ps) &  $T_\mathrm{c}$ (K) & $\Delta_\mathrm{max}$ (meV)\\
\hline
\hline
graphene (substrate) &  $0.4-1.0$ & $1.0$ & $0.1$ & \begin{tabular}{rl} $1.0$ & for $g=4$ \\  $6-35$ & for $g=2$ \end{tabular}
 & \begin{tabular}{rl} $0.1$ & for $g=4$ \\  $1-5$ & for $g=2$ \end{tabular} \\
 \hline
 graphene (suspended)         & $2.2$ & $1.0$ & 0.1 & \begin{tabular}{rl} $2$ & for $g=4$ \\  $70$ & for $g=2$ \end{tabular}
 & \begin{tabular}{rl} $0.3$ & for $g=4$ \\  $10$ & for $g=2$ \end{tabular}   \\ 
\hline
3DTI    & $0.1-1.0$ & $0.1$ & $1-10$ & $0-30$ & $0-3$  \\
\hline
2D DM with g=1 &  $0.4-2.2$ & $1.0$ & - & $50-500$ & $10-100$ \\
\hline
\hline
Cd$_3$As$_2$ DSM &  $0.1$ & $1$ & - & $0.1$ &  $0.03$ \\
\hline
TaAs WSM &  $1$ & $0.2$ & - & $2$ &  $0.3$ \\
\hline
3D DM with $g=1$ &  $1-3$ & $1$ & - & $1-20$ &  $0.3-3$ \\
\hline
3D DM with $g=2$ &  $1-3$ & $1$ & - & $10-60$ &  $1-10$ \\
\hline
3D DM with $g=4$ &  $1-3$ & $1$ & - & $1-2$ &  $0.1-0.3$ \\
\hline
\end{tabular}
\label{tab:values}
\end{table*}

 The material specific parameters of the model that control the size of the gap and $T_c$ were extracted from experimental or
 \textit{ab initio} data.
These parameters 
 are
 (i) the
dimensionless coupling constant $\alpha$, which is determined by the dielectric constant $\varepsilon$
 and the Fermi velocity $v$, (ii) the energy range for the
 Dirac states exist, i.e. the cutoff energy
scale $\Lambda$ of the effective Dirac Hamiltonian; 
this energy scale limits the range of the non-equilibrium chemical potentials $\bar{\mu}$ that can be achieved
by pumping, and (iii) the degeneracy factor $g$.
 In addition, we list the expected lifetimes $\tau$ of the transient
excitonic state inferred from experimental data for existing 2D DMs. 
 The values of the non-equilibrium chemical potentials could be estimated based on the
 density of photoexcited carriers and the properties of the laser pump in the experiment. Such analysis is
 provided for graphene, for which such experimental data is readily available in the literature.
 In other cases, we take $\bar{\mu}=\Lambda/2$.

 Below we provide a more detailed explanation for the choice of material parameters
 and discuss the results presented in Table~\ref{tab:values} for 2D and 3D DMs. We also
 discuss features of the experimental setup that could affect the detection of the transient excitonic states.
%
\subsubsection{2D DMs: graphene and topological insulators}
The effective dielectric constant for graphene is given by
$\varepsilon=(\varepsilon_\mathrm{sub}+\varepsilon_\mathrm{vac})/2$, where $\varepsilon_\mathrm{sub(vac)}$ is
the dielectric constant of the substrate(vacuum). 
Using the reported values of $\varepsilon$ for graphene on the substrate, $\varepsilon\in[2:15]$~\cite{santos2013tmd}, 
 and taking $v\approx1.0\times 10^{6}$m/s, we get $0.1\lesssim\alpha\lesssim{1.0}$. 
 For two typical substrates,  SiC and SiO$_2$, $\alpha\approx 0.4$ and $\alpha\approx 0.8$, respectively. For 
 free-standing graphene,  $\alpha\approx2.2$. 
 We consider conventional (uniform) pumping with linearly polarized light ($g=4$) 
and valley-selective pumping ($g=2$). 
In order to enable  valley-selective pumping, 
 a finite energy gap in the Dirac spectrum is required to
generate 
an orbital magnetic moment in the valleys. 
 A finite gap can be induced in graphene by inversion-symmetry breaking, e.g. due to the substrate.

We consider 3DTIs with a single Dirac cone ($g=1$),
such as Bi$_2$Se$_3$ and related materials.  
The dielectric constant in 3D TIs is taken to be $\varepsilon=(\varepsilon_\mathrm{TI}+\varepsilon_\mathrm{vac})/2$,
where $\varepsilon_\mathrm{TI}$ is the dielectric constant of the bulk. The bulk dielectric constant 
 is of the order of $100$ ($\varepsilon_\mathrm{TI}=113$ in Bi$_2$Se$_3$ and $\varepsilon_\mathrm{TI}=75-290$
in Bi$_2$Te$_3$~\cite{richter1977dielectric}), leading to $\varepsilon\approx 50$. It is possible that 
in some samples, $\varepsilon$ can be reduced to $\varepsilon\approx{30}$ due to heavy doping~\cite{Beidenkopf2011spacial}.
 Combinding the values of $\varepsilon$ and the  
 typical velocities of the Dirac states 
$v=2.0-6.0 \times 10^{5}$~m/s~\cite{pertsova2016quantum}, we obtain $\alpha$ in the range $\alpha\in[0.1:0.4]$. However, since smaller values of $v$ 
can be found e.g. in the so called ``slow'' DMs~\cite{triola2015many} and $\varepsilon$
could be tuned by gating in TI thin films, we consider a larger range $\alpha\in[0.1:1.0]$.

It can be shown that in the case of graphene, $\mu_\mathrm{e/h}=\hbar{v}\sqrt{\pi n_\mathrm{e/h}}$, 
 where $n_\mathrm{e/h}$ is the density of photoexcited carriers. 
Furthermore, carrier densities in typical experiments can be estimated based on the following relation 
$\hbar\omega_\mathrm{pump}n_\mathrm{e/h}\approx \Phi A_0$~\cite{li2012femtosecond}, 
where $\Phi$ is the pump fluence, $\hbar\omega_\mathrm{pump}$ is the pump energy, and 
  $A_0$ is the absorption coefficient of
graphene. Taking $A_0=0.02$, $\hbar\omega_\mathrm{pump}=1$~eV
and assuming $n_\mathrm{e}=n_\mathrm{h}\equiv n_\mathrm{ex}$, we 
find that a chemical potential $\bar{\mu}\approx 500$meV corresponds to carrier density 
$n_\mathrm{ex}\approx 10^{13}$~cm$^{-2}$ (and a pump fluence $\Phi\approx 100$~$\mu$Jcm$^{-2}$), 
which can be achieved in present experiments~\cite{brida2013ultrafast}.

As already mentioned in Section~\ref{pump_scheme}, 
 optical excitation in 3D TIs is different from graphene and involves bulk states, since   
 the energy of the pump pulse exceeds the insulating bulk gap 
 (the bandgap in Bi$_2$Se$_3$ is $\approx{300}$~meV while $\hbar\omega_\mathrm{pump}\approx 1.5$~eV).  
 As a result, 
 electrons are pumped from the valence band into the conduction band and then 
 scatter down to the low-energy Dirac surface states. 
 Recent trARPES experiments  showed that the lifetime of the photoexcited carriers in the upper Dirac cone 
 increases as the energy approaches the node~\cite{zhu2015ultrafast,Sumida_prb2019}. 
 In samples where the chemical potential is close to the node, the relaxation bottlence due to 
  the vanishing phase space at the node leads to long-lived population inversion~\cite{Sumida_prb2019}. 
 Based on the reported trARPES results, the non-equilibrium chemical potential in 3D TIs 
 is of the order of $100$~meV~\cite{zhu2015ultrafast}.

 For valley-pumped suspended graphene ($\alpha=2.2$, $g=2$ and $\bar{\mu}=500$~meV), we estimate 
 $\Delta_\mathrm{max}\approx 10$~meV and the corresponding $T_c\approx 70$~K, which is  
 within the resolution of typical trARPES experiments. Other experimental probes  
that may be sensitive to the presence of excitonic gaps, are time-resolved optical conductivity 
~\cite{gilbertson2011tracing} and photoluminescence measurements, which are expected to 
show a peak due to enhanced photoluminescence from recombined electron-hole pairs~\cite{versteegh2012pumpedZnO}. 
The calculation of optical conductivity in pumped 3D DMs in 
the normal and excitonic insulator
state has been carried out in Ref.~\cite{Pertsova_prb2018}.

Due to the large bulk dielectric constant, $\alpha$  is 
typically smaller in 3D TIs compared to graphene. For the minimum value  
 $\alpha= 0.1$, the estimated gap is only a fraction of a meV, i.e. 
 it is below the resolution of typical trARPES experiments. However, 
 for values of $\alpha$ closer to those in graphene on the substrate ($0.4\lesssim\alpha\lesssim{1}$), 
  a gap of few meV and $T_c$ of tens of K can obtained. Larger coupling 
  constants could be found in materials with smaller dielectric constant or smaller Fermi velocities. 
  For instance, Dirac states with $v\approx{10^4}$~m/s and 
 $\alpha\approx{7}$, were found in quasi-two-dimensional organic conductors~\cite{hirata2016organic,hirata2016organic2}.
  
  We should note that on the timescale of the population inversion in graphene ($100-200$~fs), 
 the local electronic temperature for the electron and hole distributions is still very large, e.g. it reaches few thousand 
K~\cite{gierz2013graphene, li2012femtosecond}, which greatly exceeds 
the predicted $T_c$ even in the case of suspended graphene.  
 This problem can be overcome in materials with 
 increased lifetime of the population inversion such as (Sb$_{1-x}$Bi$_x$)$_2$Te$_3$~\cite{Sumida_prb2019}, 
 where the carriers have more time to cool down. Another possibility is 
 to search for materials with desired properties that maximize $T_c$. Critical tempeartures of hundrds of K can 
 be achived in 3D TIs with a single Dirac node, a large bandgap and a relatively large $\alpha$ (see estimates
for a hypothetical 2D DM with $g=1$ and other parameters similar to graphene).

A recent experimental study reported the evidence of excitonic superfluid phase on the surface
 of Bi$_{2-x}$Sb$_x$Se$_3$ 3D TI~\cite{Hou_natcomm2019}. By using scanning photocurrent microscopy, it was shown that
 photoexcited electrons and holes form charge neutral bound states which propagate over distances
  of $1$~mm at temperatures up to 40 K.
  There are indications that the observed effect could be explained by the theoretical model discussed in this work.
 For instance, in Bi$_{2-x}$Sb$_x$Se$_3$ samples the Fermi level is close to the Dirac point which is an important condition for
 the formation of excitonic state. Moreover, the the longest lifetime of the population inversion was reported in Sb-doped 3D TIs~
 \cite{sumida2017,Sumida_prb2019}. There are no signatures of the excitonic phase in pure Bi$_3$Se$_3$, in which these conditions are
 not satisfied~\cite{Hou_natcomm2019}. Although the existence of the transient condensate with
 extremely long decay time needs to be verified further, this is a promising step
 towards realization of dynamically-induced excitonic phase.

\subsubsection{3D DMs: Dirac and Weyl semimetals}
Although many 3D DMs have been discovered recently, in Table~\ref{tab:values} two main examples are considered, Cd$_3$As$_2$ DSM and TaAs-family
of WSMs, i.e. TaAs, TaP, NbAs and NbP. For these materials, extensive ARPES data and electronic structure calculations are available, 
allowing for determination
 of the material specific
 coupling constant. Thus, for  Cd$_3$As$_2$,
$v_x\approx v_y\approx 1.3\times 10^{6}$~m/s, $v_z\approx 3.3\times 10^5$~m/s~\cite{liu_natmat2014_dsm},
$\Lambda\approx 1$~eV~\cite{liu_natmat2014_dsm}, and  $\varepsilon=36$~\cite{madelung_1998_cd2as2}.
For the TaAs family, $v\approx 2.5\times 10^5$~m/s~\cite{lee_prb2015_taas}, $\Lambda\approx 200$~meV~\cite{lee_prb2015_taas,xu2015weyl},
and $\varepsilon=10$~\cite{dadsetani_2016_taas}. Note that in the case of anisotropic Dirac dispersion, we use the average velocity
for numerical estimates. Based on these values, we estimate 
$\alpha=0.1$ and $\alpha=1$ for Cd$_3$As$_2$  and TaAs, respectively.

Typically, the degeneracy factor in realistic 3D DMs is large. Since 
screening increases with $g$, uniform pumping can not generate a sizable excitonic gap in 
these systems. 
Therefore, we  
  assume selective pumping and choose the value of $g$ that maximizes the size of the gap and $T_c$. 
  Such an optimistic estimate corresponds to $g=2$ for a WSM and $g=4$ for a DSM in the CDW phase. 
  It is possible to consider pumping on a single Dirac or Weyl node, $g=2$ and $g=1$, respectively; 
  however, in this case only the EI phase is realized leading to substantially smaller gaps.

The numerical estimates for the excitonic gap and critical temperature in 3D DMs 
are summarized in the lower half of Table~\ref{tab:values}. 
For TaAs WSM, we  predict $\Delta_\mathrm{max}\approx{0.3}$~meV and $T_c\approx{2}$K. 
Due to a small $\alpha$ resulting from a large dielectric constant, comparable to that of a 3D TI, the
 estimated gap and $T_c$ for Cd$_3$As$_2$ DSM is one order of magnitude smaller.  
 Despite the small size, the predicted excitonic gap in these two types of 3D DMs is nethertheless finite. 
 In contrast to this, the equilibrium excitonic gap is zero, since the material-specfic coupling $\alpha$ is 
  both examples is smaller than the equilibrium critical coupling.

The last three rows in Table~\ref{tab:values} correspond to hypothetical 3D DMs with improved parameters, e.g. 
 ${1}\lesssim\alpha\lesssim{3}$ and the cutoff energy scale $\Lambda$ for the Dirac/Weyl states of the order of $1$~eV. 
As before, we consider selective pumping and a few values of the degeneracy factor, $g=1$, $2$ and $4$. The largest gap and $T_c$, comparable to 
the values for selectively pumped suspended graphene, are predicted for the CDW phase with $g=2$.

\subsection{Dynamics of the order parameter}\label{sec:dynamics}
For the numerical estimates of the gap and $T_c$ presented in Table~\ref{tab:values}, we assumed a quasiequilibrium 
 state with an 
 infinitely long lifetime. In realistic systems, the population inversion is sustained over a 
 finite time range and decays towards equilibrium 
 with a single Fermi-Dirac distribution. As a result, the excitonic gap has a finite lifetime.
 
The basic condition for observation of the transient excitonic states is 
 $\tau_\mathrm{ex}<<\tau$, where  $\tau_\mathrm{ex}=\hbar/\Delta$ is the timescale for the formation of the 
 excitonic condensate and $\tau$ is the lifetime of the population inversion.  
 Assuming $\tau\approx100-200$~fs~\cite{gierz2013graphene} and $\Delta\approx{10}$~meV
in graphene~\cite{gierz2013graphene} and
$\tau\approx{10}$~ps~\cite{Sumida_prb2019} and $\Delta\approx{1}$~meV in 3D TIs, 
this condition is satisfied and therefore excitonic gaps could 
in principle be observed. 
Apart from being difficult to observe in experiment due to finite energy resolution, 
 smaller gaps would require larger lifetimes of the population inversion.

In the case of pumped graphene, which has been studied extensively, the global carrier dynamics is governed by two main processes, the  
 initial thermalization due to carrier-carrier and carrier-phonon scattering which occurs on the time scale of $50-100$~fs
 and further phonon-induced carrier cooling which happens on ps time scale~\cite{malic2011graphene}.  The population inversion, 
 if realized, develops within the first (fs) stage of the dynamics.  
It has been shown theoretically that the most important mechanism for build-up of population inversion is the 
phonon-induced intraband scattering, which leads to scattering of the highly-excited 
carriers down towards the Dirac node resulting in the filling of the states at low energies. This process is 
responsible for a carrier relaxation bottleneck close the Dirac point and leads to population inversion. 
  The  decay of the population inversion on a $100-200$~fs timescale, as observed in the experiment~\cite{li2012femtosecond}, 
  is caused by Coulomb-induced Auger recombination~\cite{malic2011graphene}. Neglecting Auger recombination processes in theoretical 
  simulations based semiconductor Bloch equations leads to a long-lived population inversion, 
  inconsistent with reported experimental values. 
  
  Despite graphene being the best candidates in terms of the size of the
gaps and $T_c$, the lifetime of population inversion in typical hole-doped graphene on the substrate is relatively short ($100-200$~fs). 
 Longer lifetimes are expected in undoped graphene with chemical potential at the Dirac node, 
 where recombination of electron-hole pairs is suppressed due to vanishing phase space at the node. 
 Continuous pumping could be used to sustain the population inversion over 
 sufficiently long timescales, effectively leading to a quasiequilibrium excitonic state. 
 However, in this scheme high temperatures of the photoexcited carriers will be distructive to 
 excitonic condensation.
  
A recent theoretical work suggested that population inversion
could be also realized in 3D DMs~\cite{Afanasiev_prb2019}. 
It was shown that ultrarelativistic electronic dispersion of WSMs
 leads to strong suppression of Auger recombination due to phase
space restrictions imposed by energy and momentum conservation. Considering that 
the overall dynamics in 3D DMs is qualitatively similar to that of graphene~\cite{Lu_prb2017_ultrafast_DSM,Weber_japl2017,Weber_apl2018}, 
  suppressed Auger recombination can result in a population inversion on the timescales that are several
orders of magnitude larger than in graphene and are comparable or larger than in 3D TIs. 
This makes pumped WSM promising candidates for transient excitonic states,
provided that other material parameters are such that screening effects are minimal.

Finally, the relaxation of the order parameter from the 
established excitonic condensate state towards equilibrium can be studied 
 using  a time-dependent approach~\cite{Triola_prb2017}, based on semiconductor Bloch equations
~\cite{lindberg1988be,haug2004quantum,stroucken2011optical,malic2011graphene,winzer2013microscopic}. 
In this approach we introduce  electron and hole populations, $n_{\textbf{k}}^\mathrm{e(h)}=\left\langle 
{c}_{\textbf{k},\mathrm{e(h)}}^{\dagger}c_{\textbf{k},\mathrm{e(h)},}\right\rangle$, and the so called interband polarization, or anomalous 
correlator, $f_{\textbf{k}}=\left\langle{c}_{-\textbf{k},\mathrm{e}}c_{
\textbf{k},\mathrm{h}}^{\dagger}\right\rangle$, which is related to the order 
parameter $\Delta_{\mathbf{k}}\equiv\Delta({\mathbf{k}})$ as $\Delta_{\textbf{k}}=\sum_{\textbf{k}^{\prime}}V_{\textbf{k}-
\textbf{k}^{\prime}}f_{\textbf{k}^{\prime}}$ [see the definition of the gap in Eq.(~\ref{eq:gap_def})]. The dynamics  
of the single-particle expectation values is governed by the following system of 
 differential equations
%
\begin{equation}
\begin{aligned}
\frac{d n_{\mathbf{k}}^\mathrm{e}}{dt}&=i\Delta_\mathbf{k}^{*}f_\mathbf{k}^{*}-i\Delta_\mathbf{k}f_\mathbf{k}
+\frac{d n_{\mathbf{k}}^\mathrm{e}}{dt}|_\mathrm{scat},\\
\frac{d n_{-\mathbf{k}}^\mathrm{h}}{dt}&=i\Delta_\mathbf{k}^{*}f_\mathbf{k}^{*}-i\Delta_\mathbf{k}f_\mathbf{k}
+\frac{d n_{\mathbf{k}}^\mathrm{h}}{dt}|_\mathrm{scat},\\
\frac{d f_{\mathbf{k}}}{dt}&=i(\varepsilon_{\mathbf{k}}^\mathrm{e}+\varepsilon_{\mathbf{k}}^\mathrm{h})f_{\mathbf{k}}
+i\Delta_\mathbf{k}^{*}(1-n_\mathbf{k}^\mathrm{e}-n_{-\mathbf{k}}^\mathrm{h})+
\frac{d f_{\mathbf{k}}}{dt}|_\mathrm{scat},
\label{SBE}
\end{aligned}
\end{equation}
%
where $d/dt|_\mathrm{scat}$ denotes the phenomenological scattering terms. 

The scattering terms can be calculated by coupling the electron and hole subsystems to 
featureless fermionic or bosonic reservoirs and by integrating 
out the reservoir degrees of freedom~\cite{stefanucci2013nonequilibrium, goldstein2015photoinduced}. 
In this simple model, we take into account interband and intraband scattering, 
characterized by the relaxation times $T_1$ and $T_1^\prime$, respectively. 
The interband scattering leads to the recombination of electron-hole 
 pairs, which results in the decay of the population inversion. Intraband scattering leads to 
  thermal equilibration of electrons/holes within the corresponding Fermi-Dirac distributions described by the 
  instantaneous chemical potentials $\mu_{\mathrm{e(h)}}(t)$ at time $t$. The interband and intraband 
   scattering terms are given by, respectively
\begin{equation}
\begin{aligned}
\frac{d n_{\mathbf{k}}^\mathrm{e(h)}}{dt}|_\mathrm{scat}&=-\frac{n_{\mathbf{k}}^\mathrm{e(h)}(t)-n_{\mathrm{F}}
(\mu_\mathrm{e(h)}(t))}{T_1^{\prime}}
-\frac{n_{\mathbf{k}}^\mathrm{e(h)}(t)}{T_1},\\
\frac{d f_{\mathbf{k}}}{dt}|_\mathrm{scat}&=-\frac{f_{\mathbf{k}}(t)}{T_2}.\label{SBE:scat}
\end{aligned}
\end{equation}
%
Note that both carrier-carrier and carrier-phonon scattering can contribute to intraband and interband processes. 
The model in Eq.~(\ref{SBE:scat}) does not differentiate between different relaxation paths but takes 
into account the overall intra- and interband relaxation with separate relaxation rates.  
Microscopic dynamical equations have been derived in the case of graphene, using 
the second-order Born-Markov approximation~\cite{malic2011graphene}.

\begin{figure*}[ht!]
\centering
\includegraphics[width=0.98\linewidth,clip=true]{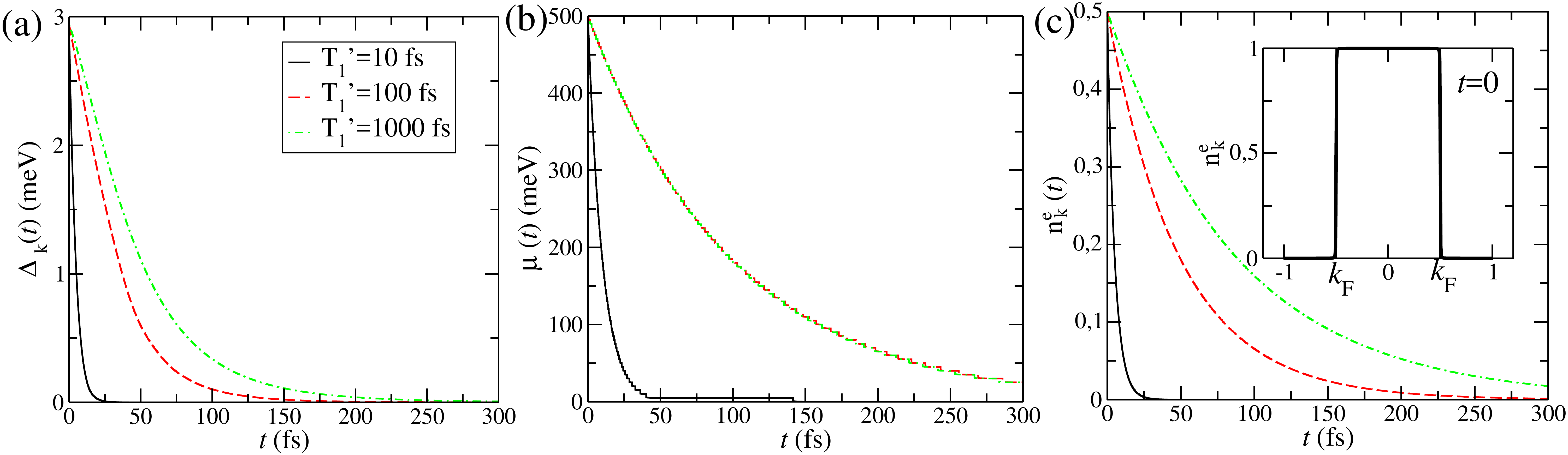}
\caption{Time evolution of (a) the order parameter, (b) the electron occupation and 
(c) the non-equilibrim chemical potential obtained from numerical 
integration of Eq.~(\ref{SBE}) with scattering terms defined in Eq.~(\ref{SBE:scat}), for a 2D DM 
 with $T_1=100$~fs and $T_1^\prime=10$, $100$ and $1000$~fs. 
 Here $g=2$, $\alpha=0.7$, $T=0$. The value of the non-equilibrium potential 
 before the relaxation ($t=0$) is $\bar{\mu}=500$~meV. $\Delta_\mathbf{k}$ and $n_\mathbf{k}^e$ are 
 calculated at $k=k_\mathrm{F}$, where $k_\mathrm{F}$ is the Fermi wavevector in equilibrium.  
The inset in panel (b) shows the distribution of $n_\mathbf{k}^\mathrm{e}$ at $t=0$. The dynamics of electrons and holes 
 are identical to each other.}
\label{fig_dyn}
\end{figure*}

The relaxation of the gap is controlled by the dephasing time $T_2$. It can 
 be shown that it is related to the interband and intraband scattering times as 
$T_2^{-1}={T_1^\prime}^{-1}+T_1^{-1}$~\cite{malic2011graphene}. 
Figure~\ref{fig_dyn} 
 shows the time-evolution of the order parameter, electron occupation and non-equilibrium 
 chemical potential for a 2D DM with parameters close to those of graphene 
 on the substrate. 
 
 As the initial state at $t=0$, we 
 take the values of the gap and electron and hole 
 occupations calculated using the self-consistent gap equation, 
 assuming the quasiequilibrium excitonic state with $\bar{\mu}=\mu_\mathrm{e}=-\mu_\mathrm{h}$.  
At later times, the excitonic gap and the 
  non-equilibrium 
  chemical potentials are calculated self-consistently at each time step, by using  Eq.(~\ref{SBE}) 
 and the definition of the gap.  
 Note that here we focus only on the relaxation of the transient excitonic state towards equilibrium. 
 Thus, we do not model the scattering processes associated with optical excitation and with building up of the 
 population inversion on the fs timescales, e.g. the interplay between carrier-phonon intraband 
 scattering and Auger recombination in the case of graphene. 
 To be consistent with the results for graphene, we fix the interband scattering time to $T_1=100$~fs and consider a 
 few values for $T_1^\prime$. The simulation time is limited to $300$~fs.
 
 Since all relaxation channels contribute to the dephasing of the order parameter, 
the lifetime of the transient excitonic 
state is determined by the shortest of the relaxation times, or the largest scattering rate. This 
is illustrated in Fig.~\ref{fig_dyn}. For the regime $T_1^\prime\gtrsim{T_1}$ ($T_1^\prime=100$~fs and 
$T_1^\prime=1000$~fs curves), the 
decay time of the non-equilibrium chemical potential $\bar{\mu}(t)$, which determines the population inversion, 
is the same and is controled by $T_1$. The relaxation
  of the gap [Fig.~\ref{fig_dyn}(a)] and the electron/hole populations 
  [Fig~\ref{fig_dyn}(b)] is slightly different in the two cases but it is also mainly goverened by $T_1$ and 
  is of the order of $100$~fs. 
  However, for the case $T_1^\prime<T_1$, the gap decays to zero within tens of fs. 

As mentioned above, 
experimental results~\cite{gierz2015graphene} and microscopic modeling~\cite{winzer2013microscopic} 
suggest that the Coulomb-induced interband interactions, in particular Auger
recombination, have the largest scattering rate and is
 the main source of decay of the population inversion on the timescale of $100-200$~fs.
Therefore, the main result of the simple dynamical model is that at least for the case of graphene, this 
gives an estimate for the lifetime of the transient excitonic
state. More detailed experimental investigations and microscopic modelling similar to that done for graphene~\cite{malic2011graphene}  
are needed to determine the scattering mechanisms and their contributions to the 
relaxation of the order parameter in other materials.
 
\section{Conclusions and outlook}\label{concl}
In summary, we have reviewed the recent proposal for transient excitonic condensate in pumped DMs in the context of
a broader search for dynamically-induced quantum states of matter. We have
compared optically pumped DMs with other promising platforms for excitonic condensation, both in equilibrium and in non-equilibrium.
The unique feature of DMs is the strongly energy-dependent DOS resulting from Dirac dispersion, which
 results in a tunable enhancement of the strength of the Coulomb interaction compared to
the values accessible in equilibrium. We have shown using examples of 2D and 3D DMs that
 there exist a range of material parameters, in which optical pumping is more advantageous for excitonic condensation
 compared to equilibrium.

 We have discussed the properties of the transient excitonic
 condensate, e.g. spectroscopic signatures and the complex phase diagrams that
result from the competition  between electronic screening effects and the enhancement of DOS by pumping.
The key features that characterize these exotic states are energy gaps in the quasiparticle spectrum which open up at the non-equilibrium chemical
potentials of photoexcited electrons and holes. Such gaps and the corresponding suppression of spectral weight near the gaps
could be detected by photoemission spectroscopy. Numerical estimates for specific materials indicate
that the largest effect (gaps of the order of $10$~meV and $T_c$ up to $70$~K) could be achieved in graphene. 3D TIs with a single Dirac
cone, in particular (Sb$_{1-x}$,Bi$_x$)$_2$Te$_3$ and similar compounds, are also promising candidates due to prolonged lifetime of the transient population 
inversion~\cite{Sumida_prb2019}.

Recently, transport signatures of the excitonic condensate phase have been found on 3D TI surfaces, with critical temperature consistent with
the values predicted by the theoretical model~\cite{Hou_natcomm2019}. Although the existence
of the condensate requires rigorous confirmation, this work gives an incentive to use 3D TIs as
a model system for further investigation of the transient excitonic states. One interesting problem is
the effect of spin-momentum locking on the spin texture of the excitonic states and their dynamics.
 Furthermore, it was pointed out in previous theoretical work
 that non-equilibrium excitonic condensate in pumped semiconductors should exhibit persistent,
 self-sustained oscillations in the induced photo current and in the dynamics of the order parameter~\cite{Ostreich1993,
 Szymaska_prl2006,Perfetto2019}. 
A fascinating question that can be addressed using the theoretical model of pumped DMs augmented by dynamical equations
for the order parameter, is whether there is an analogue of such effect in pumped semimetals with
 Dirac dispersion.

Finally, we proposed general guidelines to search for promising material candidates, which include
  (i) large coupling, which requires small Fermi velocity and small dielectric constant, 
 (ii) small Dirac cone degeneracy, which is crutial for reducing screening effects, and 
 (ii) large $\Lambda$ (the energy extent of
the Dirac states). In addition, long-lived population inversion should be possible
to allow for observation of the excitonic states on experimental timescales.
Based on these criteria, we considered several hypothetical 2D and 3D systems, in which large gaps and critical temperatures comparable to
 those found in graphene could be achieved. Given the high pace of material discovery facilitated by high-performance material
informatics, it is likely that more candidates will be identified.
Combined with continuing advances in time-resolved spectroscopies, 
 we believe that the ongoing search for transient collective states in Dirac materials will remain an active topic of research.
 
 \section{Acknowledgements}
We are grateful to G. Aeppli, P. Hofmann and Villum collaborators, H. Rostami, P. Sukhachov, O. Tjernberg,  C. Triola, J. Weissenrieder 
for uselful discussions. Work is supported by VILLUM 
FONDEN via the Centre of Excellence for Dirac Materials (Grant No.11744), 
Knut and Alice Wallenberg Foundation (Grant No. KAW 2018.0104) and the European Research Council
ERC-2018-SyG HERO.






\bibliographystyle{andp2012}
\bibliography{Driven_Dirac_new}

\end{document}